\documentclass[epj]{svjour}

\usepackage{graphics}
\usepackage[pdftex,colorlinks=true]{hyperref}

\renewcommand{\theenumi}{\roman{enumi}}
\newcommand{\pt}{$p_{\mathrm{T}}$}

\newcommand{\jpsi}{J/$\psi$}

\newcommand{\MS}{{MS}}
\newcommand{\VT}{{VT}}

\newcommand{\pp}{p^{+}}
\newcommand{\mm}{p^{-}}

\newcommand{\rhop}{\rho^{+}}
\newcommand{\rhom}{\rho^{-}}

\hypersetup{
    colorlinks = false,
    linkcolor = blue,
    anchorcolor = red,
    citecolor = blue,
    filecolor = red,
    pagecolor = red,
    urlcolor = red
}

\begin{document}




\title{
Evidence for the production of thermal muon pairs with masses above 1
GeV/$c^2$ in 158A GeV Indium-Indium collisions
}

\subtitle{NA60 Collaboration}
\author{
R.~Arnaldi\inst{11} \and
K.~Banicz\inst{4,6} \and
K.~Borer\inst{1} \and
J.~Castor\inst{5} \and 
B.~Chaurand\inst{9} \and 
W.~Chen\inst{2} \and
C.~Cical\`o\inst{3} \and  
A.~Colla\inst{11} \and 
P.~Cortese\inst{11} \and
S.~Damjanovic\inst{4,6} \and 
A.~David\inst{4,7} \and 
A.~de~Falco\inst{3} \and 
A.~Devaux\inst{5} \and 
L.~Ducroux\inst{8} \and 
H.~En'yo\inst{10} \and
J.~Fargeix\inst{5} \and
A.~Ferretti\inst{11} \and 
M.~Floris\inst{3} \and 
A.~F\"orster\inst{4} \and
P.~Force\inst{5} \and
N.~Guettet\inst{4,5} \and
A.~Guichard\inst{8} \and 
H.~Gulkanian\inst{12} \and 
J.M.~Heuser\inst{10} \and
M.~Keil\inst{4,7} \and 
L.~Kluberg\inst{9} \and 
Z.~Li\inst{2} \and
C.~Louren\c{c}o\inst{4} \and
J.~Lozano\inst{7} \and 
F.~Manso\inst{5} \and 
P.~Martins\inst{4,7} \and  
A.~Masoni\inst{3} \and
A.~Neves\inst{7} \and 
H.~Ohnishi\inst{10} \and 
C.~Oppedisano\inst{11} \and
P.~Parracho\inst{4,7} \and 
P.~Pillot\inst{8} \and 
T.~Poghosyan\inst{12} \and
G.~Puddu\inst{3} \and 
E.~Radermacher\inst{4} \and
P.~Ramalhete\inst{4,7} \and 
P.~Rosinsky\inst{4} \and 
E.~Scomparin\inst{11} \and
J.~Seixas\inst{7} \and 
S.~Serci\inst{3} \and 
R.~Shahoyan\inst{4,7} 
\thanks{\emph{Corresponding author:} ruben.shahoyan@cern.ch}
\and 
P.~Sonderegger\inst{7} \and
H.J.~Specht\inst{6} \and 
R.~Tieulent\inst{8} \and 
G.~Usai\inst{3} \and 
R.~Veenhof\inst{7} \and
H.K.~W\"ohri\inst{3,7}.
}

\institute{
Laboratory for High Energy Physics, Bern, Switzerland. \and  
BNL, Upton, New York, USA. \and                              
Universit\`a di Cagliari and INFN, Cagliari, Italy. \and     
CERN, Geneva, Switzerland. \and                              
LPC, Universit\'e Blaise Pascal and CNRS-IN2P3, Clermont-Ferrand,
  France. \and                                               
Physikalisches Institut der Universit\"{a}t Heidelberg,
  Germany. \and                                              
IST-CFTP, Lisbon, Portugal. \and                             
IPN-Lyon, Univ.\ Claude Bernard Lyon-I and CNRS-IN2P3,
  Lyon, France. \and                                         
LLR, Ecole Polytechnique and CNRS-IN2P3, Palaiseau,
  France. \and                                               
RIKEN, Wako, Saitama, Japan. \and                            
Universit\`a di Torino and INFN, Italy. \and                 
YerPhI, Yerevan, Armenia.                                    
}

\date{Date: 11/12/2008}

\abstract{
The yield of muon pairs in the invariant mass region 1$<$M$<$2.5~GeV/c$^2$ produced 
in heavy-ion collisions significantly exceeds the sum of the two expected
contributions, Drell-Yan dimuons and muon pairs from the decays of D 
meson pairs. These sources properly account for 
the dimuons produced in proton-nucleus collisions.
In this paper, we show that dimuons are also produced in excess in 158
A GeV In-In collisions. We furthermore observe, by tagging the dimuon vertices, 
that this excess is not due to enhanced D meson production,
but made of {\em  prompt} muon pairs, as expected from a 
source of thermal dimuons specific to high-energy nucleus-nucleus collisions. 
The yield of this excess increases significantly from peripheral
to central collisions, both with respect to the Drell-Yan yield and to 
the number of nucleons participating in the collisions.
Furthermore, the transverse mass distributions of the excess dimuons 
are well described by an exponential function, with inverse slope 
values around 190~MeV. The values are independent of mass and
significantly lower than those found at masses below 1~GeV/$c^2$,
rising there up to 250 MeV due to radial flow. This 
suggests the emission source of thermal dimuons above
1~GeV/$c^2$ to be of largely partonic origin, when radial flow has not
yet built up. 
}

\authorrunning{NA60 Collaboration}
\titlerunning{Evidence for thermal-like IMR dimuon production in In-In collisions.}

\maketitle

%

\section{Introduction}
\label{intro}

\begin{figure*}[htbp]
\begin{centering}
\resizebox{\textwidth}{!}{%
\includegraphics*{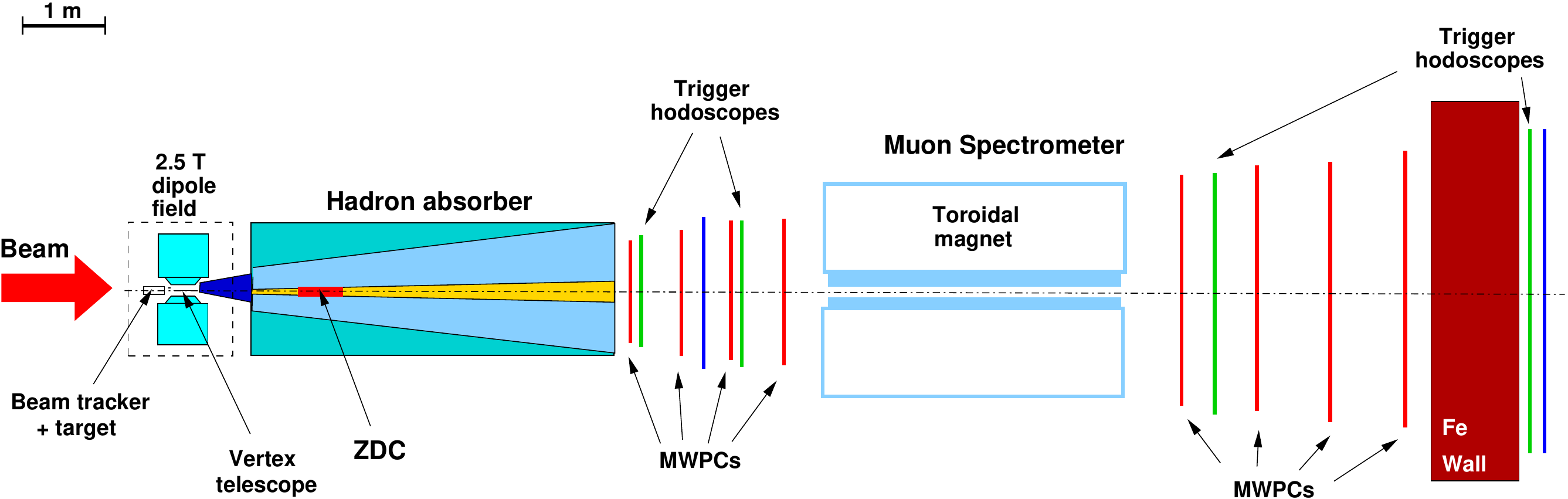}}
\caption{Schematic representation of the NA60 experimental layout. 
}
\label{fig:na60setup}
\end{centering}
\end{figure*}

%
The intermediate mass region (IMR) of the dimuon mass spectrum,
between the $\phi$ and the \jpsi\ resonances, is expected to be
well suited to search for thermal dimuon production ~\cite{SHUR80}, due to the favourable 
relative production yield with respect to the other contributions
(Drell-Yan dimuons, meson decays, etc).

Intermediate mass dimuon production in proton-nucle\-us and heavy-ion
collisions was previously investigated by NA38~\cite{NA38IMR} and
HELIOS-3~\cite{HELIOS3} in \mbox{p-W} and \mbox{S-U(W)} collisions at
200~GeV, and by NA50~\cite{NA50IMR,CSOAVE} in \mbox{p-A} (where A stands for Al,
Cu, Ag, W) and \mbox{Pb-Pb} collisions, respectively at 450 and
158~GeV.
All three experiments reported a very reasonable description of the
IMR opposite-sign dimuon mass continuum measured in the ``elementary''
proton-nucleus collisions. 
This continuum could be accounted for by a superposition of the two expected
``signal'' processes, the Drell-Yan dimuons and the muon pairs
resulting from simultaneous semi-muonic decays of (correlated) D and
$\overline{\rm D}$ mesons, on the top of a ``background'' contribution.
The latter is due to uncorrelated decays of pions and kaons 
and can be estimated from the measured like-sign muon pairs.
In contrast, all experiments observed that the opposite-sign dimuon
samples collected in the heavy-ion collision systems, taking into
account the estimated background contribution, significantly
exceeded the level expected from Drell-Yan and ``open
charm'' sources, calculated using the same procedures that
successfully reproduced the p-A data.
%

%
As discussed in~\cite{CAPELLI}, two prime interpretations were able to
describe
the findings of NA38 and NA50 equally well: the long-sought thermal
dimuons~\cite{RAPPSHUR}, and an increase in the open charm production
cross section per nucleon, from \mbox{p-A} to  \mbox{A-A}
collisions~\cite{LEVAI95}.
Several other reasons which could increase the yield of IMR 
dimuons were also proposed. Alternatively to charm enhancement,
the number of muon pairs entering the phase space
window of the experiment could be increased, while the total charm
production cross section remains unchanged. 
This could result from, \textit{e.g.}, the smearing of
the D/$\overline{\rm D}$ pair correlation resulting from rescattering 
of the charmed quarks or mesons in the surrounding dense matter~\cite{LIN98}.
%
It was also suggested
that the mass spectrum of the Drell-Yan
dimuons could be modified in heavy-ion collisions with respect to the
proton-nucleus case, because of higher-twist effects that increase the
yield of low mass dimuons~\cite{Qiu}.  Another source of dimuons that
could be present in heavy-ion collisions, while being negligible in
p-nucleus collisions, is secondary Drell-Yan production, where the
quark-antiquark annihilation uses valence antiquarks from produced
pions~\cite{Spieles}.

A decisive step in understanding the origin of the excess dimuons is
to clarify the decade-long ambiguity between prompt
dimuons and off-vertex muon pairs. In the first case they
can be thermal dimuons or extra Drell-Yan dimuons; in the second case they result from decays of D mesons, which
have a relatively long lifetime: $c\tau = 312$~$\mu$m for the $\rm
D^{+}$ and 123~$\mu$m for the $\rm D^{0}$.
The clarification of the physical origin of the IMR dimuon excess was
one of the main motivations of the NA60 experiment.  Thanks to its
ability to measure the {\it offset} of the muons with respect to the
interaction vertex, NA60 can separate, on a statistical basis, the
prompt dimuons from the off-vertex muon pairs.

In this paper we present a study of the intermediate mass dimuons
produced in In-In collisions at 158~GeV/nuc\-leon, 
based on data collected by the NA60 experiment in 2003.  The
paper is organized as follows: Section~\ref{sec:setup} describes the
NA60 experimental setup, the data reconstruction procedure and the
general performance of the apparatus; Section~\ref{sec:bgsub} explains
in some detail the background subtraction procedure;
Section~\ref{sec:result} presents the results.
Preliminary results were presented before~\cite{RSQM07}.

\section{The NA60 experimental setup and data reconstruction}
\label{sec:setup}

\begin{figure}[ht!]
\centering
\resizebox{0.99\columnwidth}{!}{%
\includegraphics*{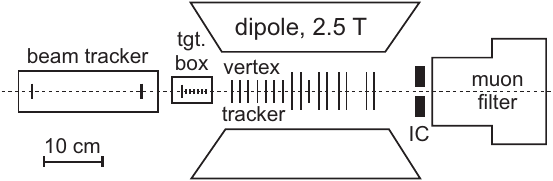}}
\caption{View of the target region. From the left: 2 stations of the
   beam tracker followed by 7 Indium targets and 16 planes of the
   vertex tracker.
}
\label{fig:tgtregion}
\end{figure}

Figure~\ref{fig:na60setup} shows a general view of the NA60 apparatus.
Its main components are the muon spectrometer (MS), previously used by
the NA38 and NA50 experiments~\cite{MSNA60}, and a novel
radiation-hard silicon pixel vertex tracker (VT)
with high granularity and high readout speed~\cite{refVT}, 
placed inside a 2.5~T dipole magnet just downstream
of the targets (see Fig.~\ref{fig:tgtregion}).  
%
The first spectrometer is separated from the second by
a hadron absorber with a total effective
thickness of $\sim$\,14~$\lambda_{\rm int}$ and $\sim$\,50~$X_{0}$.

Before entering into more details, we list the key features of this
somewhat unique setup:
\begin{itemize}
\item The vertex telescope tracks all charged particles {\it upstream}
of the hadron absorber
and determines their momenta independently of the MS, free from 
multiple scattering effects and energy loss fluctuations in the
absorber. The matching of the muon tracks before and after the
absorber, both in {\it coordinate and momentum space}, strongly
improves the dimuon mass resolution in the low-mass region (less so
higher up), significantly reduces the combinatorial background due to
$\pi$ and K decays and makes it possible to measure the muon
offset with respect to the primary interaction vertex.
\item The additional bend by the dipole field in the target region
  deflects muons with lower momenta into the acceptance of the MS,
  thereby strongly enhancing the opposite-sign dimuon acceptance, in
  particular at low masses and low transverse momenta, with respect to
  all previous dimuon experiments. A complete acceptance map in
  two-dimensional M-$p_T$ space is contained in~\cite{SANJAR2}.
\item The selective dimuon trigger and the 
  radiation-hard vertex tracker with its high read-out speed allow the
  experiment to run at very high rates for extended periods,
  maintaining the original high luminosity of dimuon experiments
  despite the addition of an open spectrometer.
\end{itemize}
%

A detailed description of the muon spectrometer can be found
in~\cite{MSNA60}. 
Its magnet defines the rapidity window 
where the dimuons are accepted, $3<y_{\rm lab}<4$.  
The trigger system is based  on four hodoscopes along the beam
direction. 
Each of them has hexagonal symmetry, with the sextants operating
independently. 
The system provides a highly selective
dimuon trigger requiring that the four hodoscope slabs hit by 
each muon match one of the predefined patterns. This ensures that the muon 
was produced in the target region.  In addition, the trigger imposes 
that the two muons must be detected in different sextants.

A detailed description of 
the radiation-hard silicon pixel vertex tracker used by NA60 in 2003 can be found 
in~\cite{ADThese}.
The readout pixel chips,
developed 
for the ALICE and
LHC-B experiments~\cite{PIXREF}, operate with a 10~MHz clock
frequency.  The VT can provide up to 12 space points per track, 9
of them from planes oriented such that the horizontal coordinate (in
the bending plane) is measured with higher precision.

The target system is composed of seven Indium sub-targets, 1.5~mm
thick each and separated by an inter-distance of 8~mm, adding up to an
interaction probability of 16\,\% for the incident Indium ions.
%
An interaction counter is located downstream of the VT. It is made of 2 
scintillator blades, appropriately holed to let the beam pass through. They
are independently read by 2 photomultiplier tubes in coincidence and thus 
allow to tag the interactions in the target region.
%
%
A beam tracker~\cite{BSRef}, made of two tracking stations 20~cm apart, 
was placed upstream of the target system.
The beam tracker allows to 
measure the flight path of the incoming ions and to derive the
transverse coordinates of the interaction point, in the target, with
an accuracy of 20~$\mu$m, 
independently of the collision centrality.
%
%
A zero degree calorimeter, previously used in NA50, is located in the
beam line, inside the muon filter, just upstream of the Uranium beam
dump.  It 
estimates the centrality of each
nucleus-nucleus collision through the measurement of the energy
deposited by the beam ``spectator'' nucleons.

Around 230 million dimuon triggers were 
recorded
on tape during the
2003 run.  In approximately half of these events a dimuon was
reconstructed from the muon spectrometer data.  Two data samples were
collected, with different currents in the ACM toroidal magnet: 4000~A
and 6500~A.  The higher field reduces the acceptance in the highly
populated region of low transverse mass ($m_{\rm T}$) muon pairs,
thereby increasing the number of high mass dimuon events collected per
day, for a constant lifetime of the data acquisition system.
The details of the event selection can be found in ~\cite{ADThese}.

The reconstruction of the raw data proceeds in several steps.  First,
muon tracks are determined from the data of the eight MWPCs and
validated by the hits recorded in the trigger hodoscopes.  If the
event has two reconstructed muon tracks which fulfil the trigger
conditions, the 
tracks in the silicon planes of the vertex tracker are
also reconstructed, and the interaction vertices are searched
for.  The track reconstruction efficiency is
$\sim$\,95\,\% for peripheral In-In collisions and $\sim$\,90\,\% for
the most central ones.  The key step in the data reconstruction
is the matching between the muon track, extrapolated from
the muon spectrometer to the target region, and the charged tracks
found in the vertex tracker.  This is done by selecting those
associations between the MS tracks and the VT tracks
which give the smallest weighted squared distance ({\it matching} $\chi^2$)
between these two tracks, in the space of angles and inverse momenta,
taking into account their error matrices.
%
The matching procedure
combines the good MS momentum resolution ($\sigma_{p}/p \sim 2$\,\%)
with the excellent VT angular precision ($\simeq$\,1~mrad) to obtain
the kinematics of the muon before undergoing the multiple scattering
and energy loss induced by the hadron absorber.  This
procedure, as mentioned before, improves the dimuon mass resolution,
from 70-80~MeV/$c^2$ to 20-25~MeV/$c^2$ in the $\omega$ and $\phi$ mass
region, and allows to correlate the muon's trajectory with the
interaction vertex, the point of primary interest for the present paper.

\begin{figure}[htbp]
\centering
\resizebox{0.95\columnwidth}{!}{%
\includegraphics*{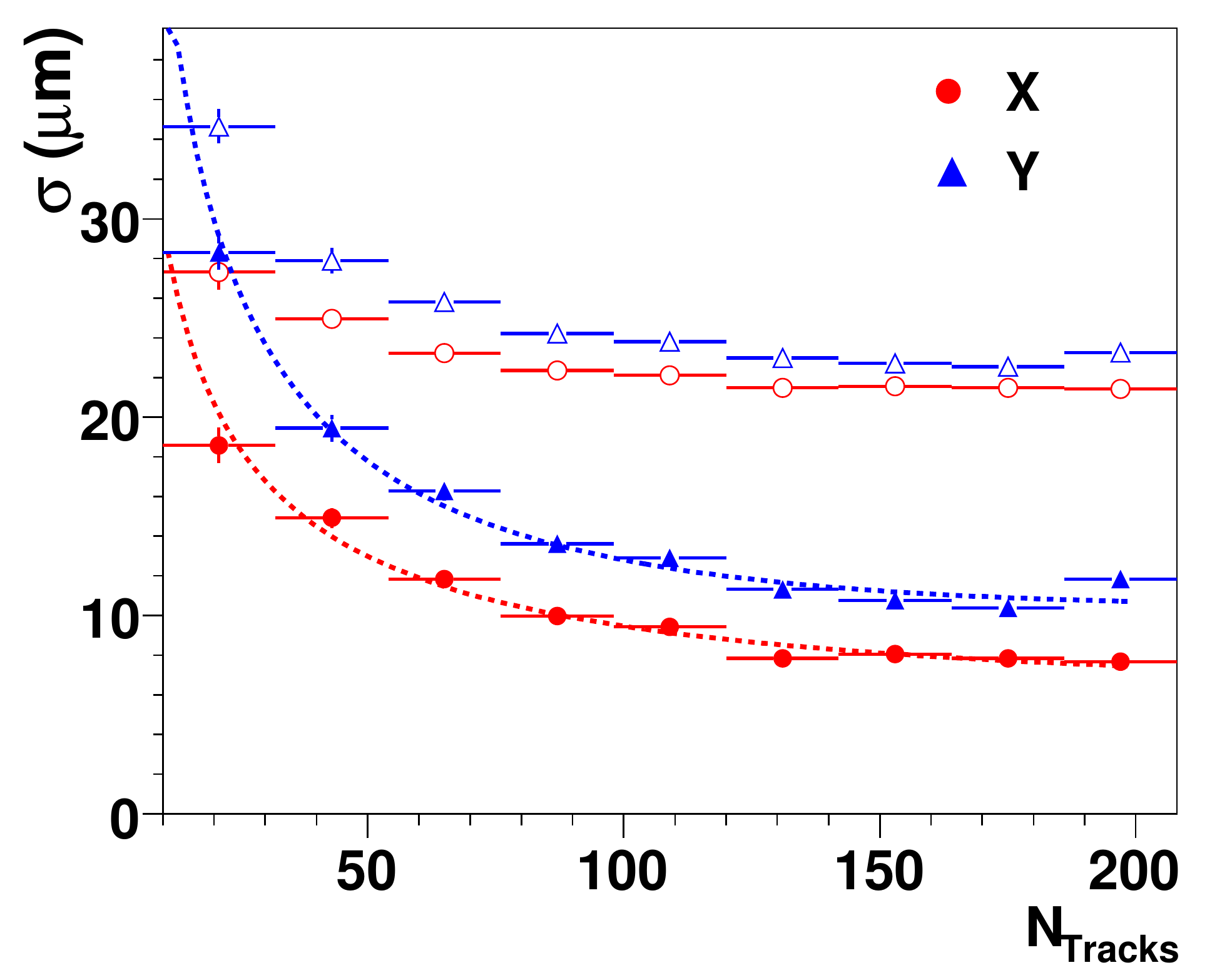}}
\caption{Dispersion between the transverse coordinates of the interaction vertex
     given by backtracing the VT tracks and by extrapolating the beam
     tracker measurement (open symbols), as a function of the
     number of tracks attached to the vertex. Derived vertex
     resolution (solid symbols).}
\label{fig:vtbs}
\vglue -2mm
\end{figure}

The resolution of the vertex determination, and its dependence on the 
number of tracks associated with the vertex, can be obtained from the
dispersion between the measurements provided by the beam tracker and by the
vertex tracker.  As shown in Fig.~\ref{fig:vtbs}, it is better than 10~$\mu$m in $x$ 
and 15~$\mu$m in $y$, except for the most peripheral collisions.

\begin{figure}[htbp]
\centering
\resizebox{0.99\columnwidth}{!}{%
\includegraphics*{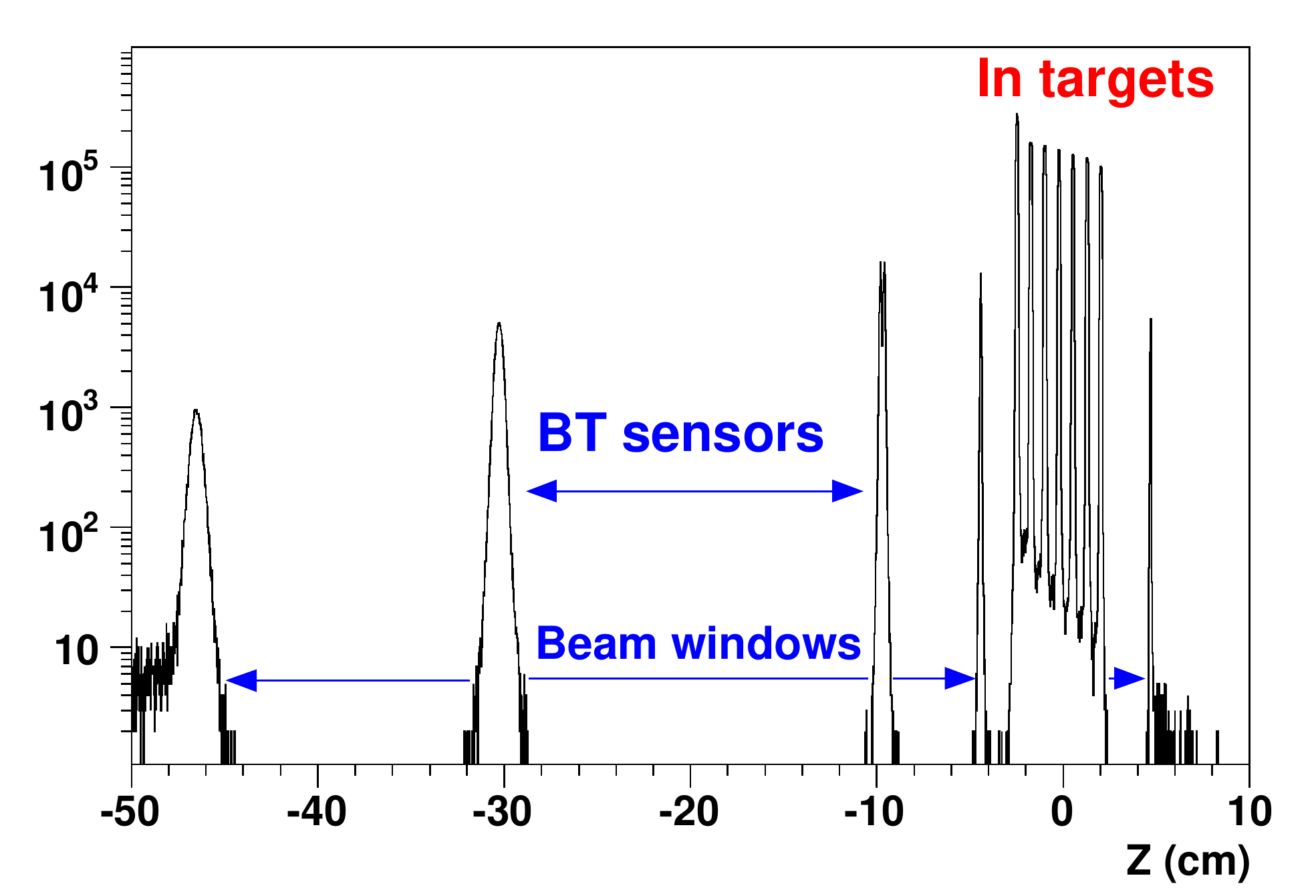}}
\caption{Distribution of the $z$ coordinate of the interaction vertex
for the events that only have one good vertex reconstructed.}
\label{fig:zv}
\vglue -2mm
\end{figure}

Figure~\ref{fig:zv} shows the distribution of the $z$ coordinate of
the reconstructed vertices, for the events with only one vertex found.
We can clearly identify the seven In
targets, placed between the two windows that keep the target box in
vacuum, downstream of the two beam tracker stations, also placed in vacuum.

\begin{figure}[htbp]
\centering
\resizebox{0.9\columnwidth}{!}{%
\includegraphics*{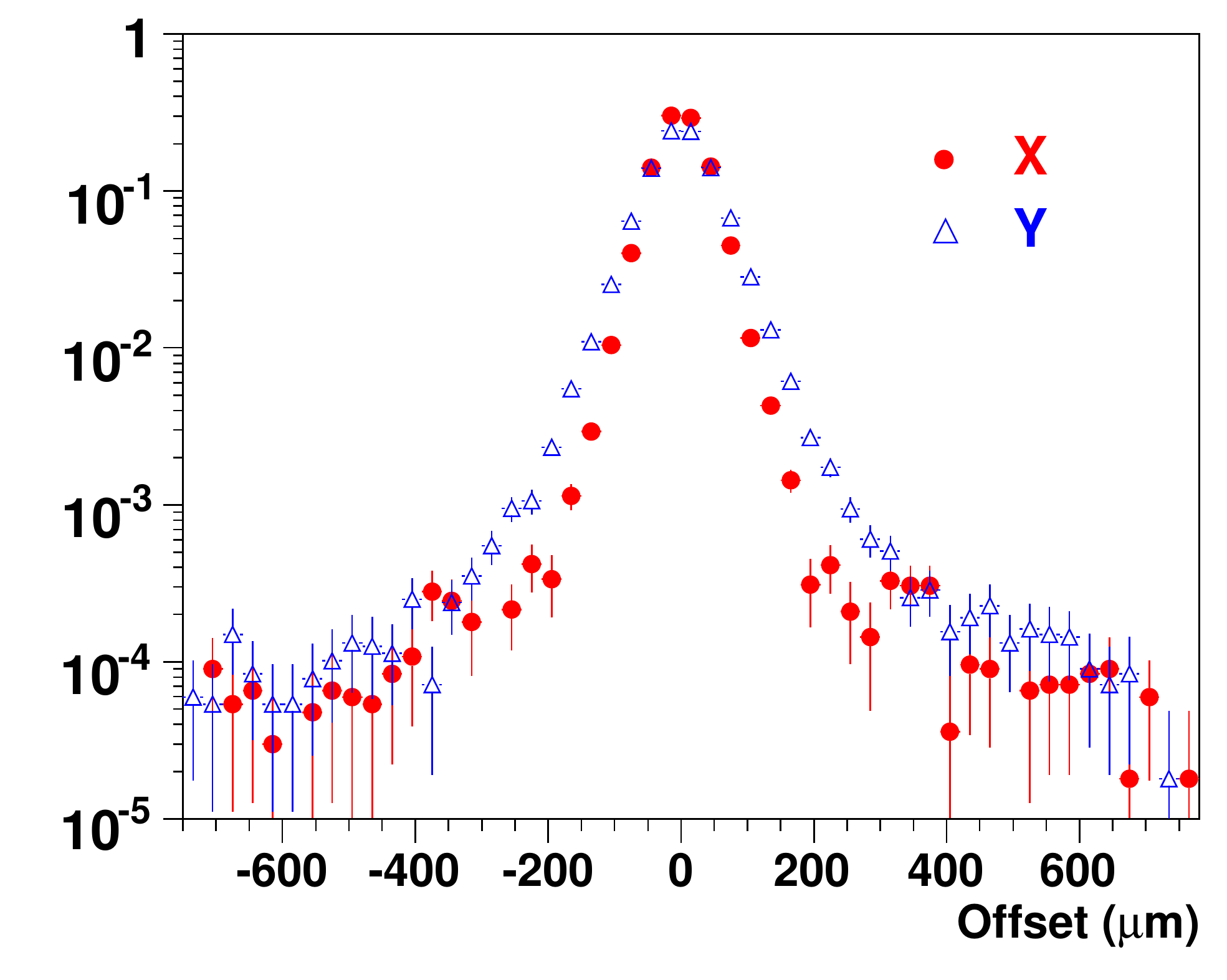}}
\caption{Offset distribution for matched muons from \jpsi\ decays, in the $x$
(circles) and $y$ (triangles) coordinates, normalized to unit area.}
\label{fig:offsres}
\vglue -2mm
\end{figure}

The resolution of the offset distance measured between
the matched muon tracks and the collision vertex can be evaluated by using the muons
from \jpsi\ decays. 
Since the $B\rightarrow {\rm J}/\psi$ contribution is negligible at SPS energies, all the \jpsi\ mesons
are promptly produced.  Moreover, the pion and kaon decay background
is negligible under the \jpsi\ peak.  Therefore, the offset
distribution made with these muons (shown in Fig.~\ref{fig:offsres})
directly reflects the resolution of the muon offset measurement: 37~$\mu$m in
$x$ (bending plane) and 45~$\mu$m in $y$.
These values are the convolution of the track
uncertainties with the accuracy of the transverse
coordinates of the vertex.

Since the offset resolution of the matched tracks is affected by
multiple scattering in the silicon planes, the analysis is performed
using a {\it weighted} muon offset variable, essentially insensitive
to the particle's momentum.  Its definition is
\begin{equation}
\Delta_{\mu} = \sqrt{\Delta x^2 V_{xx}^{-1} + \Delta y^2 V_{yy}^{-1} +
2\Delta x \Delta y V_{xy}^{-1}} \quad ,
\end{equation}
\noindent
where $V^{-1}$ is the inverse error matrix accounting for the
uncertainties of the vertex fit and of the muon kinematics fit. 
$\Delta x$ and $\Delta y$ are the differences between the 
coordinates of the vertex and those of the
extrapolated muon track, in the transverse plane crossing the beam
axis at $z=z_{\rm vertex}$.
We then characterize the dimuon through the {\it weighted dimuon
offset}, defined as
\vglue -3mm
\begin{equation}
\label{eq_offsdimu}
\Delta_{\mu \mu} = \sqrt{ (\Delta_{\mu 1}^{2} + \Delta_{\mu 2}^{2})/2}
\quad.
\end{equation}

\section{Background subtraction}
\label{sec:bgsub}

The sample of collected dimuons include a 
\textit{combinatorial background} (CB)
originating from decays of uncorrelated hadrons, mostly $\pi$'s and K's.
Since the production and decay of the parent hadrons are independent
processes, this \textit{combinatorial background} (CB)
contributes in the same way to the opposite-sign and like-sign samples
of muon pairs, increasing quadratically
with the charged particle multiplicity.  It is important to emphasise
that the NA60 trigger treats in exactly the same way the opposite-sign
and the like-sign muon pairs.

While this kind of background was already present in previous dimuon
experiments, such as NA38 and NA50, a new type of background appears
in the NA60 dimuon spectra due to the matching procedure.  Indeed, any VT
track having a small enough \textit{matching} $\chi^2$ with respect to
the muon track is considered a \textit{matching candidate}.
For high enough charged particle multiplicities, the muon track will have several
possible matches and, naturally, at most one of them is correct.
All the other associations are \textit{fake matches} and, if selected, need
to be subtracted.  Pairs where one muon match is fake and the other
one is correct are also part of the \textit{fake matches dimuon
background} (FB).  
Because the probability of having a fake
match, for each muon, is proportional to the track density, 
the yield of matched \emph{muon pairs} where only one muon is fake rises
linearly with the charged particle
multiplicity, while the yield of doubly fake pairs rises quadratically.

\begin{figure}[htbp]
\centering
\resizebox{0.99\columnwidth}{!}{%
\includegraphics*{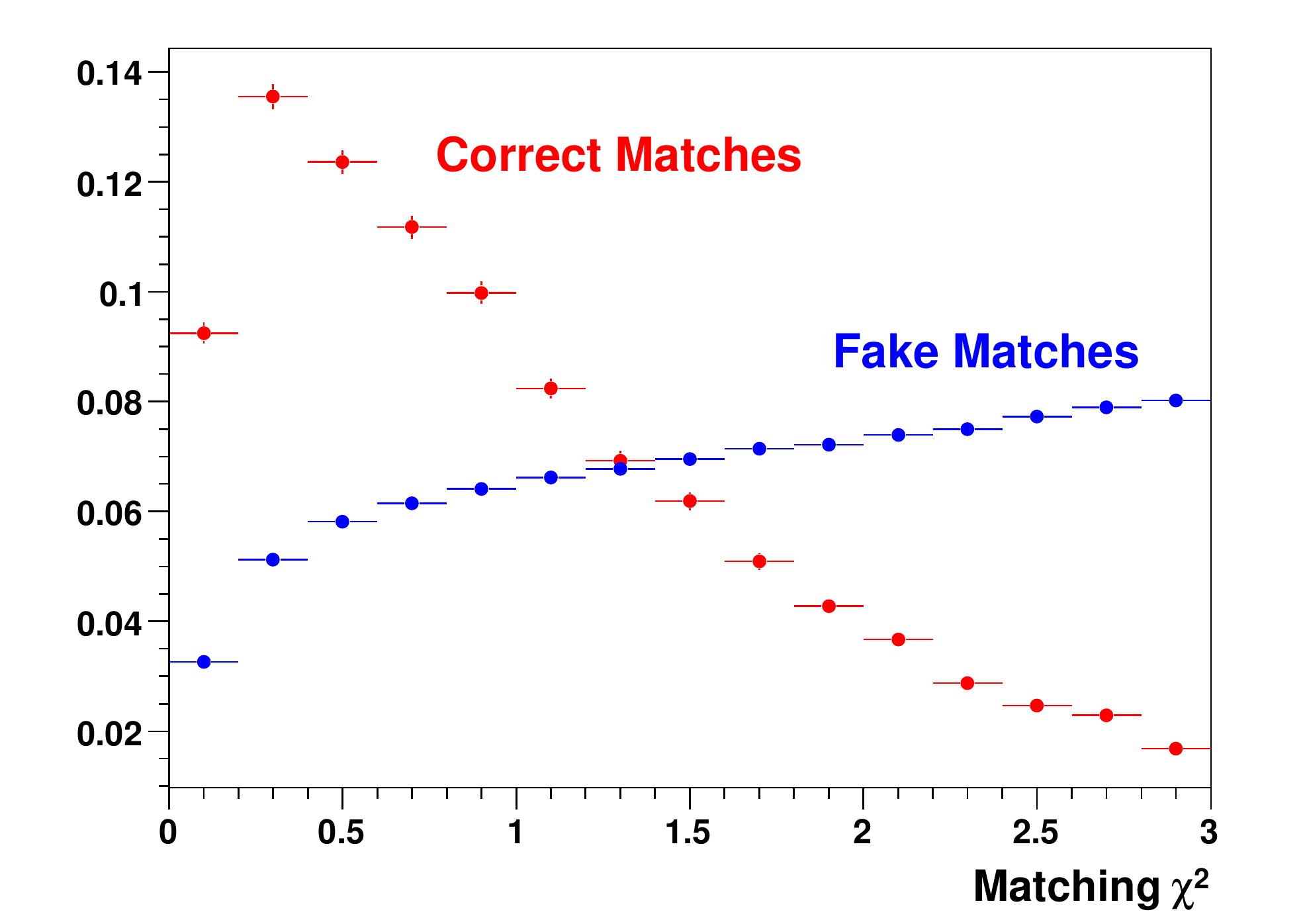}}
\caption{Matching $\chi^2$ distributions for
correct and fake matches, normalized to unit area.}
\label{fig:chi2match}
\vglue -2mm
\end{figure}

Fig.~\ref{fig:chi2match} shows the 
single muon 
matching $\chi^2$
distributions for correct and fake matches, normalized to unit
area. The distribution for fake matches is obtained from the real
data using the \textit{mixed-events technique} (see
section~\ref{sec:fakebg}):
each muon from the MS is matched with the VT tracks of another event
with the same vertex position and multiplicity. Since the probability
of having a fake match is determined only by the density of the
non-muon tracks in the phase space of the matching parameters, the
obtained distribution of the fake matches is automatically obtained
with correct normalization. 
By subtracting it from all matches, the
distribution for the correct ones is obtained.
%
One can see that the distribution for the fake matches is much flatter
than the 
corresponding distribution for
the correct matches.  Selecting exclusively matches with a matching
$\chi^2$ below a certain threshold value, 
the signal-to-background ratio can be improved at the expense of losing some
fraction of the signal.
\vglue -3mm

Despite generating the fake matches background, the muon matching
procedure leads to a very significant reduction of the combinatorial
background due to muons from pion and kaon decays.  This happens for
two reasons.  First, when the pions or kaons decay within the vertex
tracker, the kink at the decay point prevents most of the muon tracks
from being reconstructed by the silicon tracker, thereby removing these decay
muons from the matched muons sample.  Second, if
the pions or kaons decay downstream of the vertex tracker, the
matching between the meson track and the muon track usually results in
rather large matching $\chi^2$ values due to the kink between the
parent meson and decay muon. By only selecting matches of
relatively low $\chi^2$ values, we suppress the yield of
\emph{correctly matched} muons from $\pi$,K decays.


\subsection{Combinatorial background}
\label{sec:combbg}

The combinatorial background contribution to the oppo\-site-sign dimuon
distributions can be evaluated from the measured like-sign muon pair
samples, using an \textit{event mixing technique}.  Two muons from
different events with similar characteristics
are randomly picked and paired to
build the ``mixed CB'' sample. Only muon pairs respecting
the dimuon trigger conditions are kept (in particular, the mu\-ons must
be in different sextants). 
Additionally, a sextant-dependent weight is applied to the measured muons to
correct the bias introduced by the ``different-sextants'' trigger
requirement.
The mixing is done separately in 40 narrow bins in centrality in order to avoid
the bias due to the variation of the single muon kinematics and the
$\mu_{+}/\mu_{-}$ ratio within the bin width.
The technical details are given in Appendix~{\bf A}.
It is important to notice that this mixed-event procedure reproduces not only the shape of
the CB spectra but also their absolute normalization.

An important issue to consider in the CB generation concerns the selection
of the muons used for the mixing and for the computation of the
sextant-dependent weights: they must share the same target, the same field
polarities, the same charged track multiplicity bin, and the same
configuration of the silicon pixel planes (same acceptances, efficiencies, etc).
Since we are interested in determining the background contributions to
the \emph{matched} opposite-sign dimuon spectra, it could seem natural to only use for the
event mixing single muons which have at least one match.  This
would ensure that the normalizations, given by Eq.~(\ref{eqcbmixnorm}), would
be correctly computed for the matched dimuons spectra.
Unfortunately, in the case of the analysis presented in this paper,
the event mixing must be made using \textit{all} muons (including non-matched
muons), for reasons explained in the next section.  Therefore, the computed
normalizations correspond to all the dimuons reconstructed in the muon
spectrometer, matched and non-matched.


\begin{figure}[htbp]
\centering
\resizebox{0.9\columnwidth}{!}{%
\includegraphics*{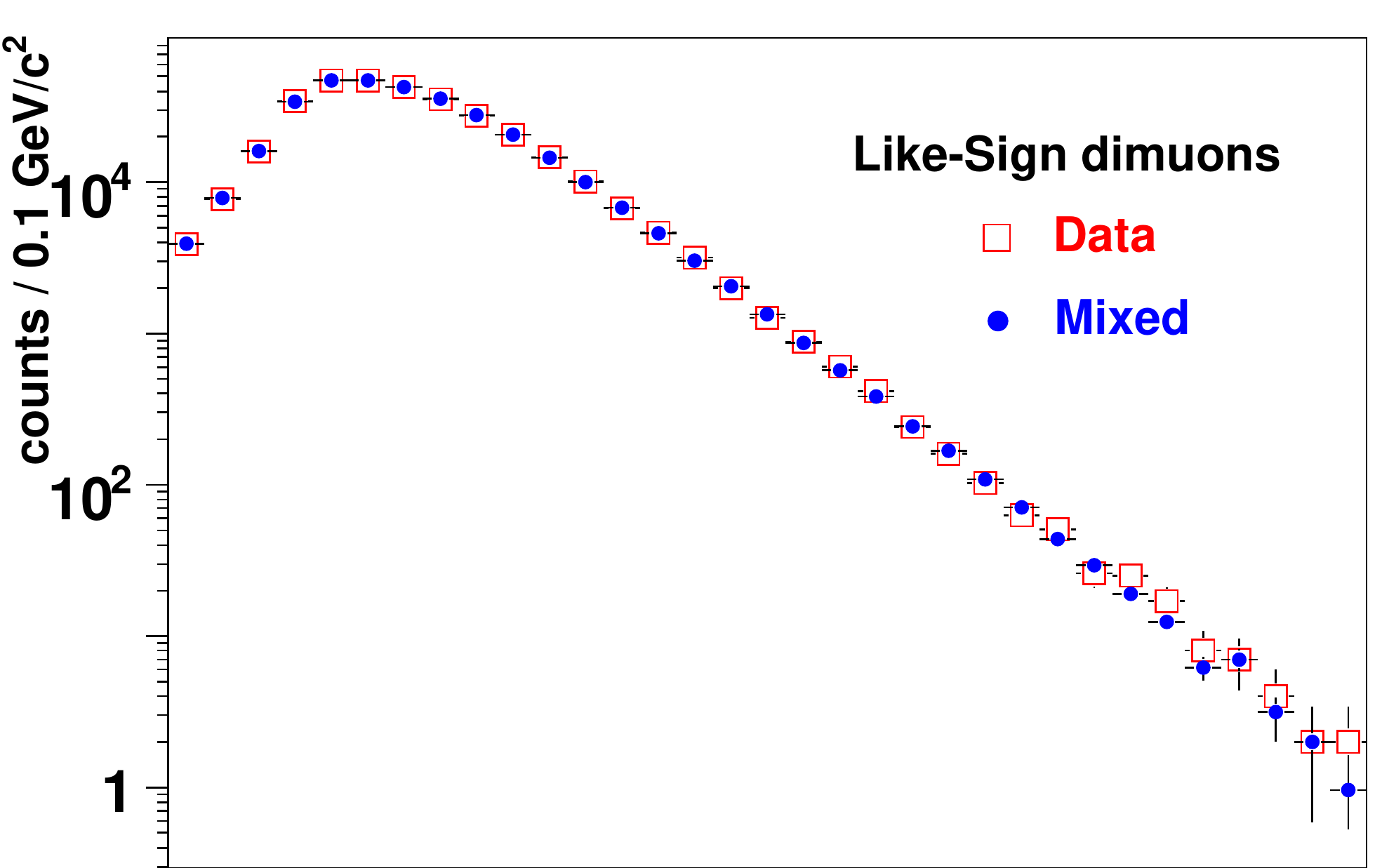}}
\resizebox{0.9\columnwidth}{!}{%
\includegraphics*{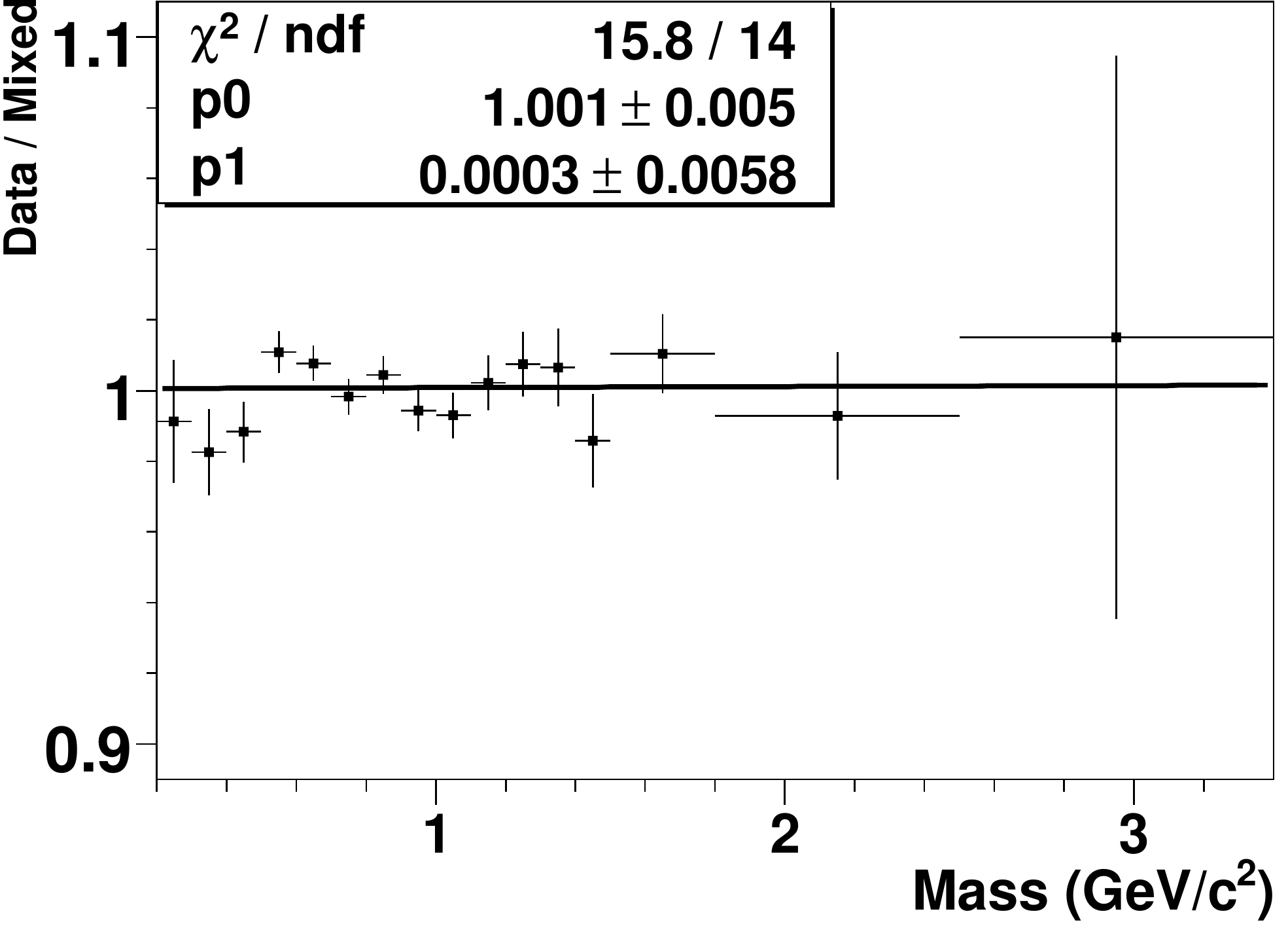}}
\caption{Measured (squares) and mixed (circles) like-sign dimuon
mass spectra (\textit{top}) and their ratio (\textit{bottom}).}
\label{fig:lsratM}
\vglue -1mm
\end{figure}

The accuracy of the mixed-event method can be
evaluated by comparing the mixed and measured distributions of
like-sign muon pairs, as shown in Figs.~\ref{fig:lsratM} and~\ref{fig:lsratD},
respectively for the mass and weighted offset (see also section~\ref{sec:bgsyserr}
below).

\begin{figure}[ht]
\centering
\resizebox{0.9\columnwidth}{!}{\includegraphics*{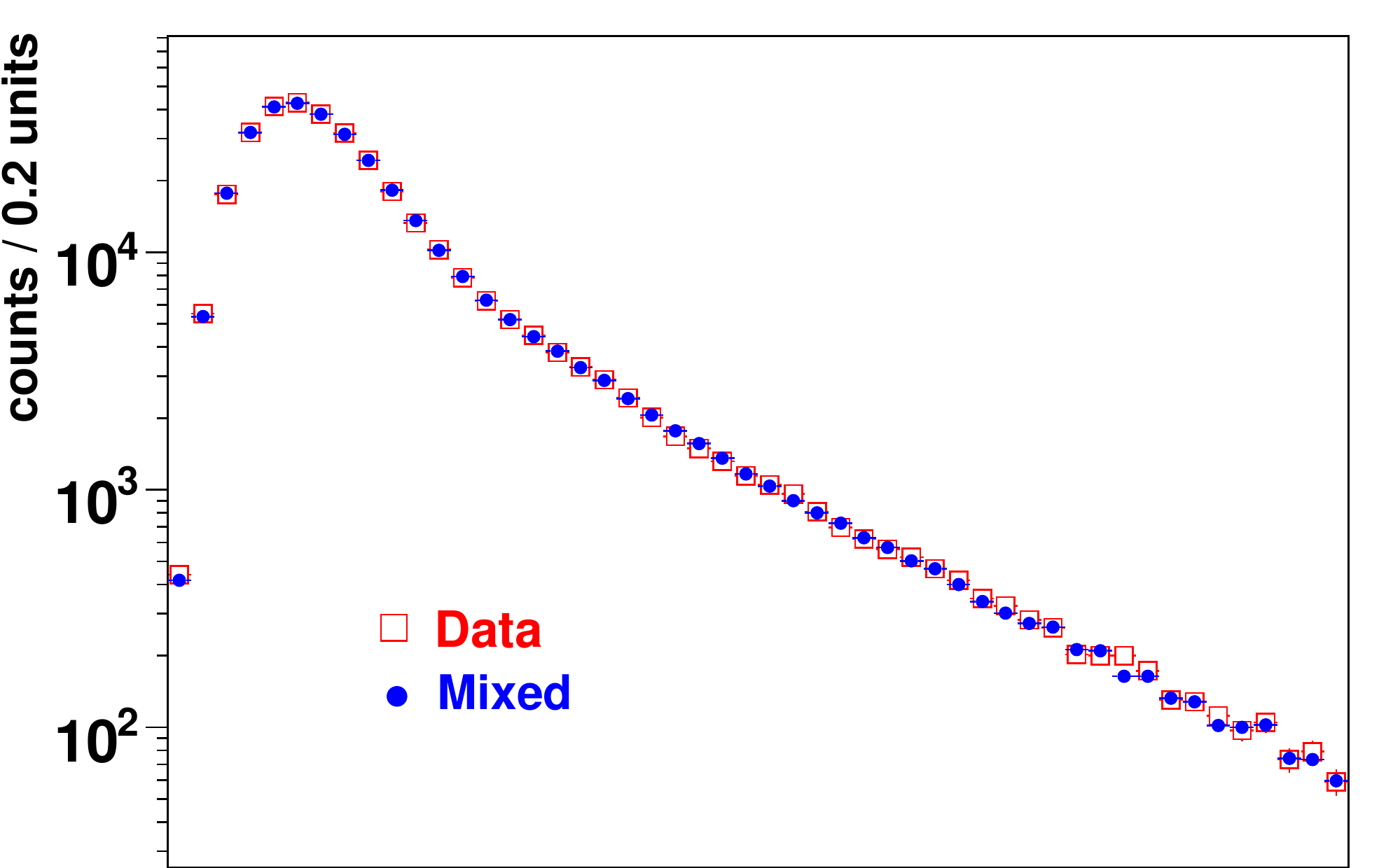}}
\resizebox{0.9\columnwidth}{!}{\includegraphics*{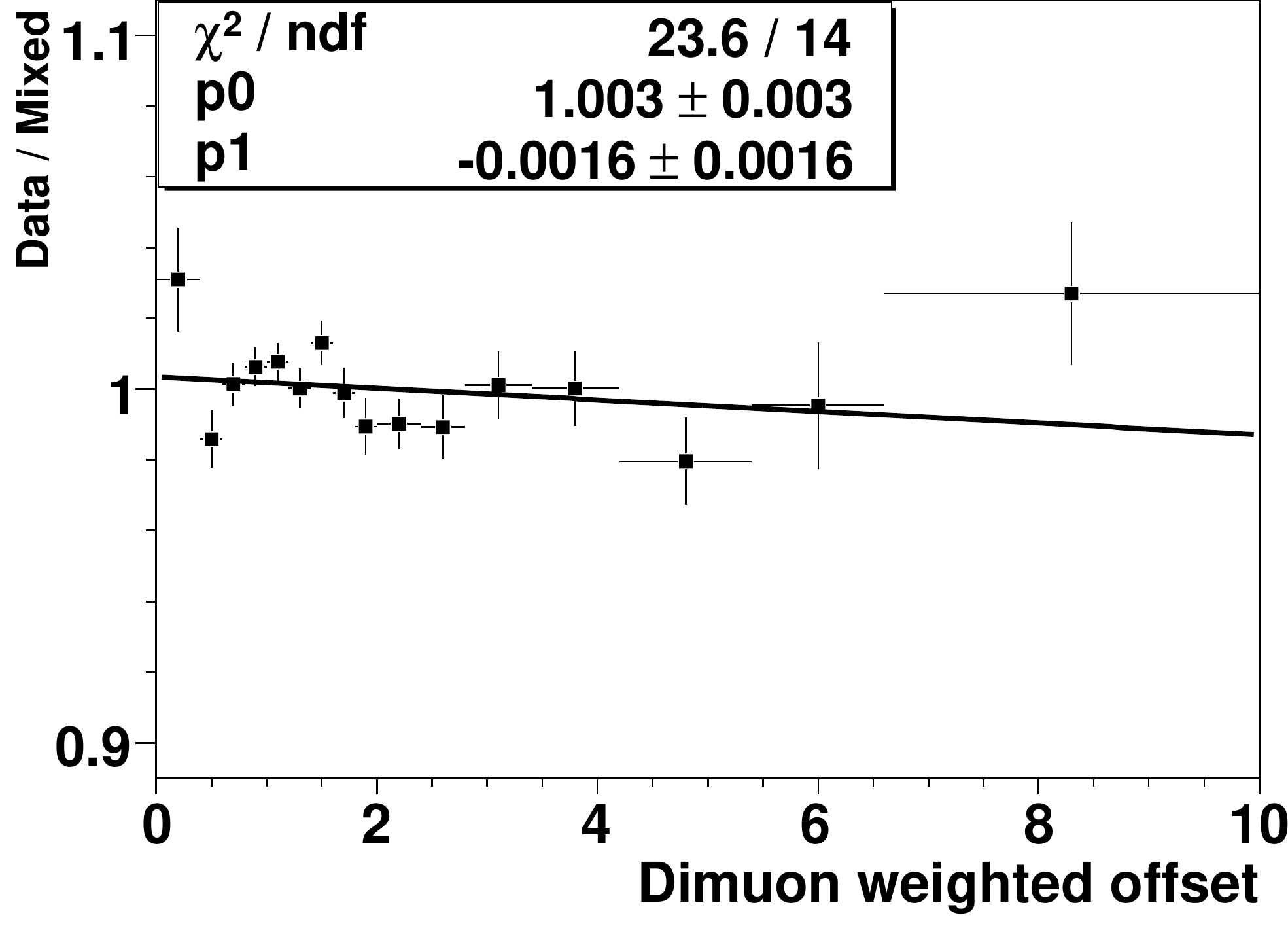}}
\caption{Same as previous figure but for the dimuon weighted offset.}
\label{fig:lsratD}
\vglue -2mm
\end{figure}

\subsection{Fake matches background}
\label{sec:fakebg}

The most direct way to evaluate the fake matches background is to use
the \textit{overlay Monte Carlo} method: the VT hits of generated
dimuons are superimposed on the VT hits of measured data,
in order to ensure realistic occupancy conditions.  By
definition, every matched track built without the sufficient
number of the Monte Carlo muon hits is a \textit{fake match} 
(conventionally we define the match as \textit{fake} if the track 
has more than two non-muon hits).  
This method works well for studies that do not
use the muon offset information.
However, while the kinematic variables of the dimuons are quite robust
with respect to the unavoidable differences between the measured and simulated
data (in particular the residual misalignment of the geometrical setup),
the offset distribution is much more sensitive.  Therefore, for studies which
require the muon offset information, such as the study reported in this paper,
we had to develop a fake subtraction procedure which only
uses measured data and provides reliable spectra in any dimuon parameter.

This is again achieved by an \textit{event mixing technique}, matching the muons of a
given event with the VT tracks of other events (with the same production target,
charged multiplicity, running conditions, etc.), for the real data
and for the artificial CB sample.  Since the real data sample also contains dimuons where
only one of the two muons was incorrectly matched, we added to our
simulated FB sample a certain proportion of muon pairs where
one of the muons retains its own match (see Appendix~B).

In principle, there could be
two alternative approaches for reproducing the matched dimuons.  If
the two muon tracks from the MS have $m_1$ and $m_2$
matches, then there are $m_1 \times m_2$ matched dimuons. 
Given the different shapes of the $\chi^{2}$ distributions for correct 
and fake matches (Fig.~\ref{fig:chi2match}), the pair composed by the two
\textit{best matches} (those with the smallest $\chi^{2}$) has the
highest probability of being the correct one. However, this
probability decreases with the increase of the number of matching
candidates, leading to a degradation of the correct matching
efficiency from peripheral to central collisions. Also, 
as explained in Appendix~B,
since it appears impossible to estimate analytically the amount of FB in such
\textit{best matches} spectra, one should rely on the
\textit{overlay Monte Carlo} method.  
Alternatively, one can consider
all $m_1 \times m_2$ matches for a given MS dimuon. Although
this will increase the amount of FB to subtract -- hence
increasing the statistical error -- such ``\textit{all matches} dimuon
spectra'' offer a better control of the systematic errors, and allow us
to cross-check the Monte Carlo based subtraction method.
The results presented in this paper are obtained using this ``all matches'' 
procedure.  Finally, 
to reproduce the offsets of the fake matches at the interaction
vertex, we apply the same algorithm as used for the CB (see Appendix~B).

The top panel of Fig.~\ref{fig:omega_fakes} shows the mass
distribution of Monte Carlo $\omega$ dimuons (with ``all matches'') and those of
the fake contributions, both as estimated by Monte 
Carlo tagging and by event mixing.  The bottom panel shows the ratio between the
mixed and the MC-tagged fake backgrounds. Both methods are seen to
agree in the mass range having meaningful statistics.
\begin{figure}[ht]
\centering
\resizebox{0.95\columnwidth}{!}{%
\includegraphics*{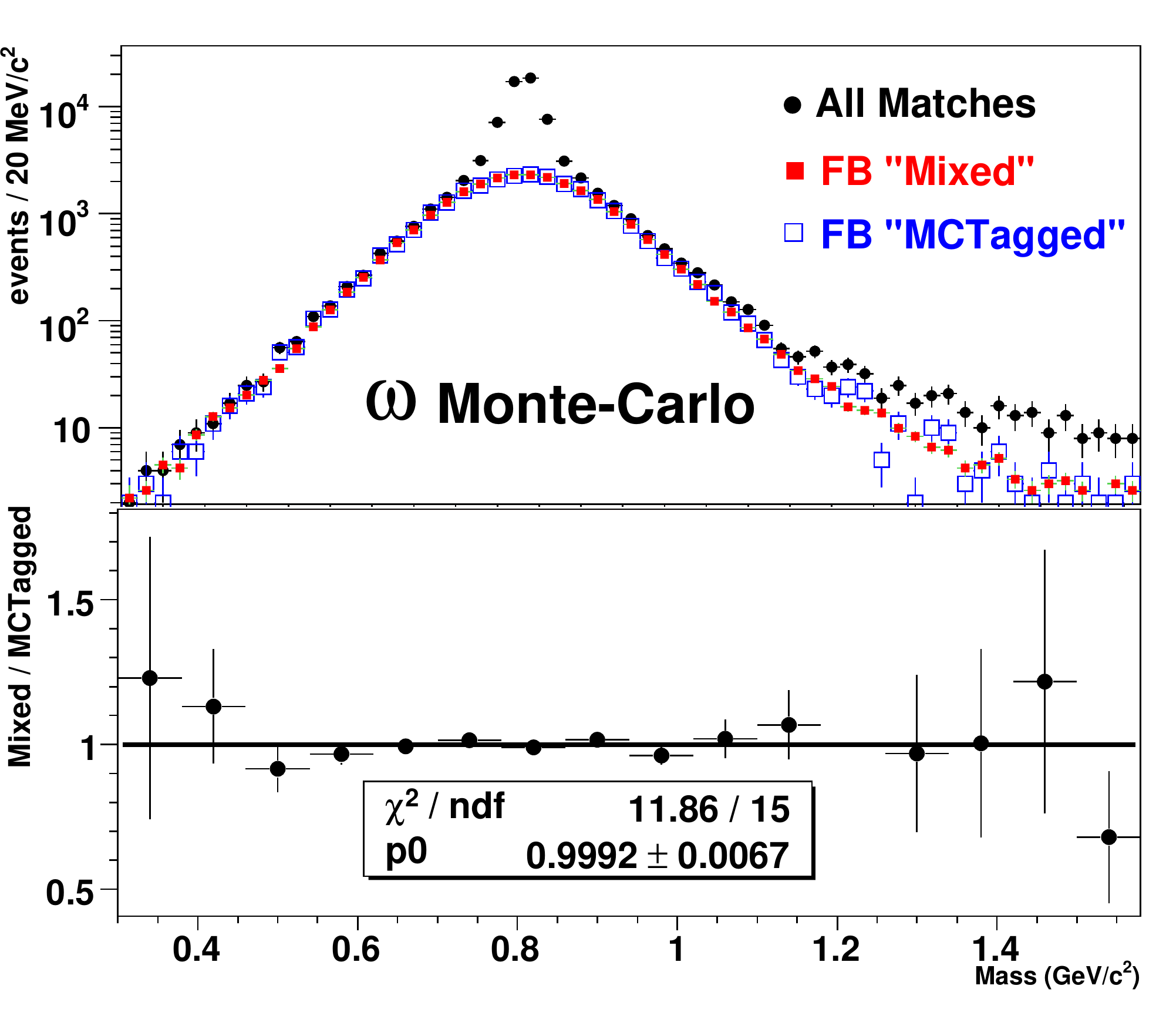}}
\caption{\textit{Top:} Dimuon mass distributions for the $\omega$
simulation, using ``all matches'' (points), and for the FB,
obtained with the \textit{event mixing technique} (filled squares)
or by Monte Carlo tagging (open squares). \textit{Bottom:} Ratio between ``mixed''
and ``MC-tagged'' fake distributions.}
\label{fig:omega_fakes}
\vglue -2mm
\end{figure}

\subsection{Extraction of the correctly matched signal}
\label{sec:signal}

The CB and FB contributions obtained via \textit{event
mixing} are not independent from each other.  As explained
above, the combinatorial pairs can be correct
or fake matches, CB = CB$_{\rm corr}$ + CB$_{\rm fake}$.
Clearly, the latter is also present in the total sample of ``fake'' pairs,
due to signal or to combinatorial dimuons,
FB = FB$_{\rm sig}$ + FB$_{\rm cb}$, with FB$_{\rm cb}$
$\equiv$ CB$_{\rm fake}$. 
Therefore, if we subtract the total fake background and the total combinatorial
background from the data, we do not obtain the correctly matched signal, because
the ``fake combinatorial'' pairs are subtracted twice:
$\rm Data - FB - CB = Data - FB_{\rm sig} - CB_{\rm corr} - 2 CB_{\rm fake} =
Signal - CB_{\rm fake}$.
To avoid this double subtraction, we must estimate the ``fake
combinatorials'' contribution.
This is done by applying the 
CB estimation algorithm described in Appendix~{\bf A} to the generated FB sample, using its
like-sign pairs for the event mixing.

\begin{figure}[b]
\centering
\resizebox{0.99\columnwidth}{!}{%
\includegraphics*{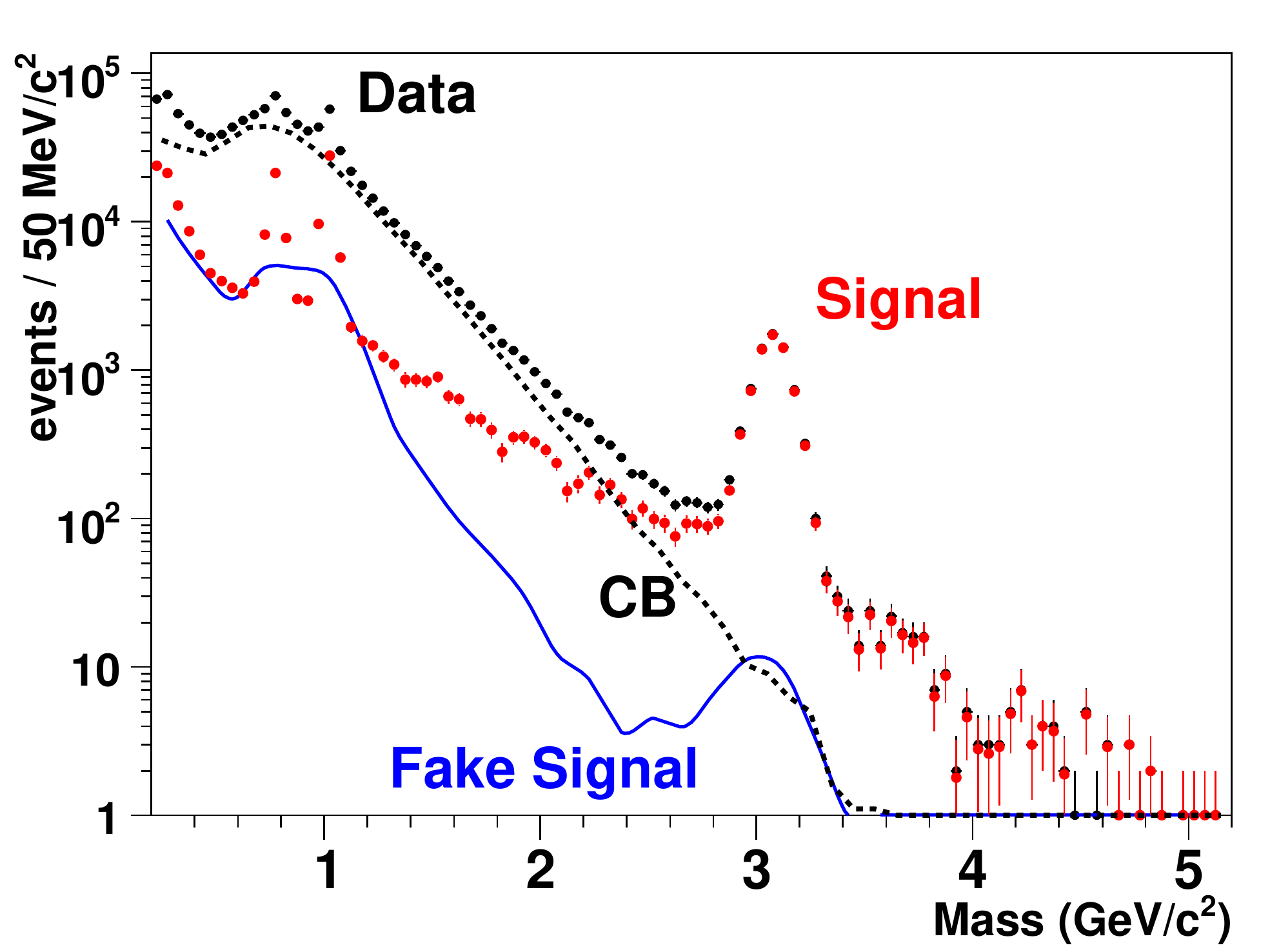}}
\vglue-2mm
\caption{The measured opposite-sign dimuon mass spectrum (points), the signal
(bullets: correctly matched; solid line: incorrectly matched), and the
combinatorial background (both correctly and incorrectly matched: 
dashed line) for the 4000~A data with \textit{matching} $\chi^2<$1.5.}
\label{fig:fullspec}
\end{figure}

Figure~\ref{fig:fullspec} shows the mass spectra (integrated over all
collision centralities) for the measured opposite-sign di\-mu\-ons, full
combinatorial and fake signal background sources, and the extracted 
correctly matched signal.


\begin{figure}[bp]
\begin{centering}
\resizebox{0.99\columnwidth}{!}{%
\includegraphics*{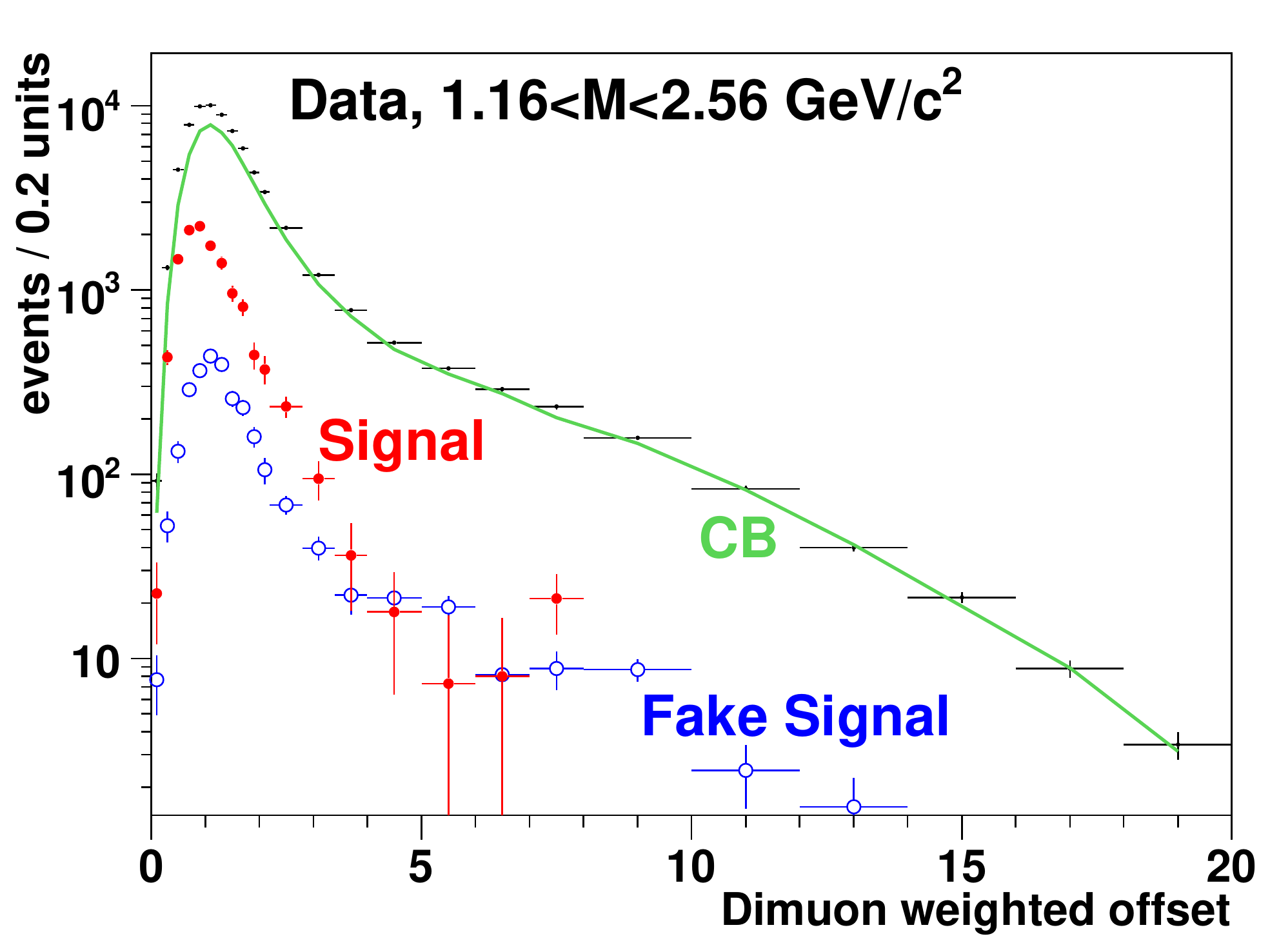}}
\vglue -0.1 cm
\caption{Data, signal and background spectra for weighted dimuon
offset distributions (defined in Eq.~\ref{eq_offsdimu}) in IMR 
for 4000~A data with \textit{matching} $\chi^2<$1.5.}
\label{fig:fullspecoffs}
\end{centering}
\vglue -0.1 cm
\end{figure}

Figure~\ref{fig:fullspecoffs} shows the \textit{dimuon weighted
offsets} distributions.  It is worth noticing that 
the region $2<\Delta_{\mu \mu}<6$,
decisive to disentangle the open charm contribution from the
prompt one, has a significant level of incorrectly matched signal.

\subsection{Systematic errors from the background subtraction}
\label{sec:bgsyserr}
\vglue -0.1 cm
The yield of like-sign dimuons remaining after the subtraction of the
mixed-event spectra constitutes a very good estimate of the residual
combinatorial background contribution left in the opposite-sign spectra.
Based on the ratios shown in Fig.~\ref{fig:lsratM},~\ref{fig:lsratD},
the accuracy of the generated background is estimated to be
$\sim$\,1\%. The resulting systematic uncertainty of the extracted
signal is defined by the signal-to-background ratio which strongly
depends on the kinematic bin and the cut imposed on the \textit{matching} $\chi^2$.
Being comparable to the statistical error, 
it changes from $\sim$\,25\% for $p_T<$0.25 GeV/$c$ at masses around
1.2~GeV/$c^2$ to $\sim$\,1\% near 2.5~GeV/$c^2$ with a lose cut
$\chi^2$=3, and it improves by more than a factor 2 for a $\chi^2$=1.5 cut.


To account for these systematic errors in the several fits mentioned 
in the next section, whenever necessary,
the \textit{statistical} errors of the estimated background were
globally scaled up to ensure that the residuals of the
like-sign spectra are compatible with zero within three standard
deviations.  Even in the worst case (when fitting a \pt\ distribution in a
narrow dimuon mass range), the errors were not increased by more 
than $\sim$\,10\,\%.

The results presented in the next section were checked by repeating
the fits with both cuts on the $\chi^2$ and both low and high magnetic
field data and were found to be always compatible within the quoted errors.

\section{Results}
\label{sec:result}

\subsection{Expected sources of IMR dimuons}
\label{sec:ExpSrc}
The analysis was separately performed for the 4000~A and 6500~A event
samples.
Dimuons were selected in the kinematical domain
defined by
$0<y_{\rm cms}<1$ and $|\cos\theta_{\rm CS}|<0.5$, where $\theta_{\rm CS}$ is
the Collins-Soper decay
angle.   
Only events with a single reconstructed vertex in one of the
seven Indium targets were kept for the analysis.
Since the signal-to-background
ratio strongly depends on the matching $\chi^2$ cut (see
Fig.~\ref{fig:chi2match}), the data were analyzed with two different cuts ($\chi^2<1.5$ and
$\chi^2<3$) to evaluate possible systematic effects related to the background
subtraction.

The shapes of the Drell-Yan and open charm contributions
were obtained with the Pythia Monte Carlo event
generator~\cite{PYTHIA} version~6.325, using the CTEQ6L set of parton
distribution functions~\cite{CTEQ} including nuclear effects through
the EKS98 model~\cite{EKS98}.  The primordial $k_{\rm T}$ was
generated from a Gaussian distribution, of width 0.8~GeV/$c$ 
for Drell-Yan (to describe the $p_{\rm T}$ distribution of dimuons heavier than
the \jpsi) and 1.0~GeV/$c$ for the open charm~\cite{charmCLHW}.  
The charm quark mass was kept at the default value of 1.5~GeV/$c^2$ 
and the $P_V$ parameter (probability that the $c$-quark hadronizes in vector
states) was set to 0.6~\cite{ADavid}, resulting in an effective $c\bar{c} \rightarrow
\mu^{+} \mu^{-}$ branching ratio 0.84\,\%.

\begin{figure}[b]
\centering
\vglue -0.5cm 
\resizebox{0.9\columnwidth}{!}{%
\includegraphics*{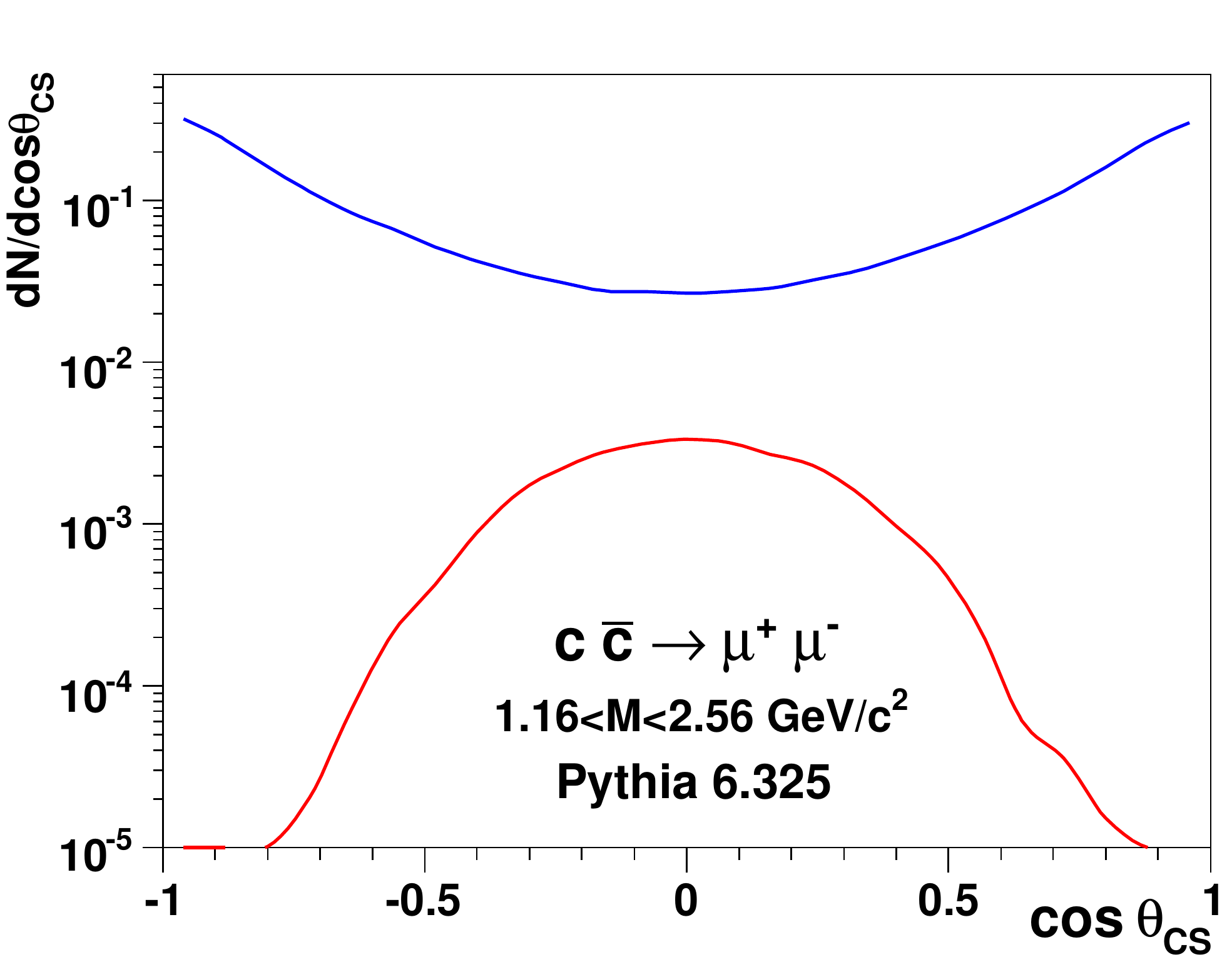}}
\vglue -0.2cm 
\caption{$\cos\theta_{\rm CS}$ distribution of $c\bar{c} \rightarrow \mu^{+} \mu^{-}$ 
dimuons in the 1.16--2.56~GeV/$c^{2}$ range, as generated by Pythia,
before (upper line, normalized to unit area) and after 
(lower line) applying the muon reconstruction algorithm.}
\label{fig:costet_charm}
\end{figure}

The Pythia events were generated at the vertices of real events and
propagated through the experimental setup using the Geant~3.2
transport code~\cite{GEANT}.  The resulting hits were added to
those of the real event used to set the interaction vertex and the resulting
\textit{overlay Monte Carlo} events were then reconstructed, with the codes used to
process the measured data.  We only kept the events surviving the  
selection cuts also applied to the real data.
It is worth noticing that the $|\cos\theta_{\rm CS}| < 0.5$ window 
significantly cuts the open charm contribution, because Pythia's strong 
D/$\overline {\rm D}$ pair correlations give a $\cos \theta_{\rm CS}$ 
distribution peaked at $-1$ and $+1$.
This means that the fraction of accepted dimuons from D pair decays is
small and very sensitive to the kinematic distributions and correlations
used in Pythia.
Figure~\ref{fig:costet_charm} shows the $\cos\theta_{\rm CS}$ distribution
of the $c\bar{c} \rightarrow \mu^{+} \mu^{-}$ dimuons, with mass in the 
range 1.16--2.56~GeV/$c^{2}$, before and after the reconstruction step.

The basic goal of the analysis is to compare the measured yield of signal IMR dimuons
to the \emph{expected} yield, from the Drell-Yan and open charm contributions.
Since this comparison is to be done as a function of collision centrality, the events are 
distributed in 12 sub-samples, using the number of tracks reconstructed 
in the VT as centrality estimator.  The corresponding average number of nucleons 
participating in the In-In collision, $N_{\rm part}$, can be derived
within the Wounded Nucleon Model  
$N_{\rm ch} \approx q N_{\rm part}$, where $N_{\rm ch}$ is the number of charged
particles produced in the VT angular window, corrected for acceptances and
efficiencies~\cite{NA60Mult}. The proportionality constant $q$ is
found by matching the observed $dN/dN_{\rm ch}$ to $dN/dN_{\rm part}$
generated from the Glauber model.

To judge if the sum of the expected contributions reproduces or not, in amplitude
and shape, the measured mass and weighted offset IMR dimuon signal distributions, 
we must determine the \emph{normalizations} of the Drell-Yan and open charm spectra.
We start by fixing the \emph{relative} normalization of the open charm contribution
with respect to the Drell-Yan contribution, by calculating the ratio of their production 
cross sections.  In order to calculate this ratio, we use the high-mass Drell-Yan cross sections
measured by NA3~\cite{NA3DY} and by NA50~\cite{NA50DY}, which show that 
Pythia's Drell-Yan spectrum needs to be up-scaled by a K-factor of 1.9, and we use a
charm cross section of $\sigma_{c\bar{c}} = 8.6$~$\mu$b.  The latter number
corresponds to the value required to reproduce the dimuon mass distributions
measured by NA50 in p-A collisions at 450~GeV~\cite{NA50IMR,CSOAVE}, 
$36\pm 3.5~\mu$b, scaled down in energy (to 158~GeV) using Pythia.  It is worth
noting that this value is around 1.8 times higher than what would be obtained
from a ``world average'' estimate~\cite{charmCLHW} based on measurements of
D meson production from fully reconstructed hadronic decays.
One should stress that neither NA50 nor NA60 are
suited to measure the full phase space $c\bar{c}$ cross section: as
can be seen from Fig.~\ref{fig:costet_charm}, the acceptance window
$|\cos\theta_{\rm CS}|<0.5$ defined by the muon spectrometer contains
less than 20\,\% of all dimuons from $D\bar{D}$ decays, and the
extrapolation of the measurement to full phase space strongly depends
on how well the correlations between the two decay muons are described
by Pythia. 
The obtained open charm to Drell-Yan cross-section ratio is then kept the same for 
all centrality bins, since both processes are expected to scale with centrality in exactly 
the same way (proportionally to the number of binary nucleon-nucleon collisions).

The next step is to determine the Drell-Yan normalization from the yield of high-mass 
dimuons.  Given the relatively small statistics, especially 
when the events are subdivided in several centrality bins, we calculate the Drell-Yan
yield from the much larger number of  \jpsi\ events and expected
\jpsi/$DY$ ratio. The latter is measured in proton-nucleus
collisions~\cite{GONCALO} by NA50 (at 400--450~GeV and scaled to
158~GeV) and is corrected for the  \jpsi\ ``anomalous suppression'' measured in In-In collisions~\cite{psiNA60}.
A 10\,\% relative systematic error is applied to the resulting normalizations, to account 
for uncertainties in the \jpsi\ anomalous suppression and normal absorption patterns.

\subsection{Data versus expectations in dimuon mass and offset}
\label{sec:datafits}

\begin{figure}[t]
\centering
\resizebox{0.97\columnwidth}{!}{%
\includegraphics*{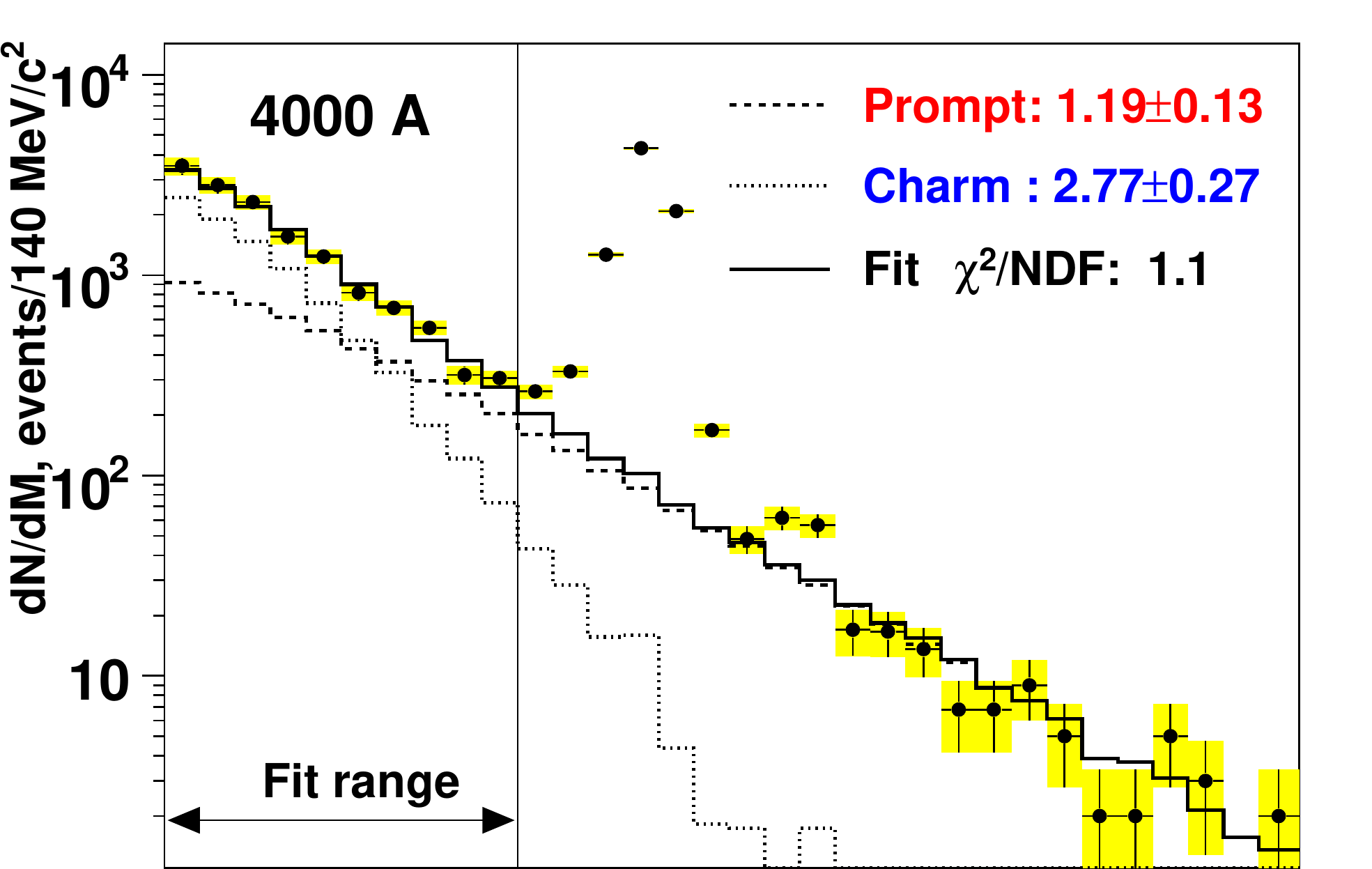}}
\vglue -2.7mm
\resizebox{0.97\columnwidth}{!}{%
\includegraphics*{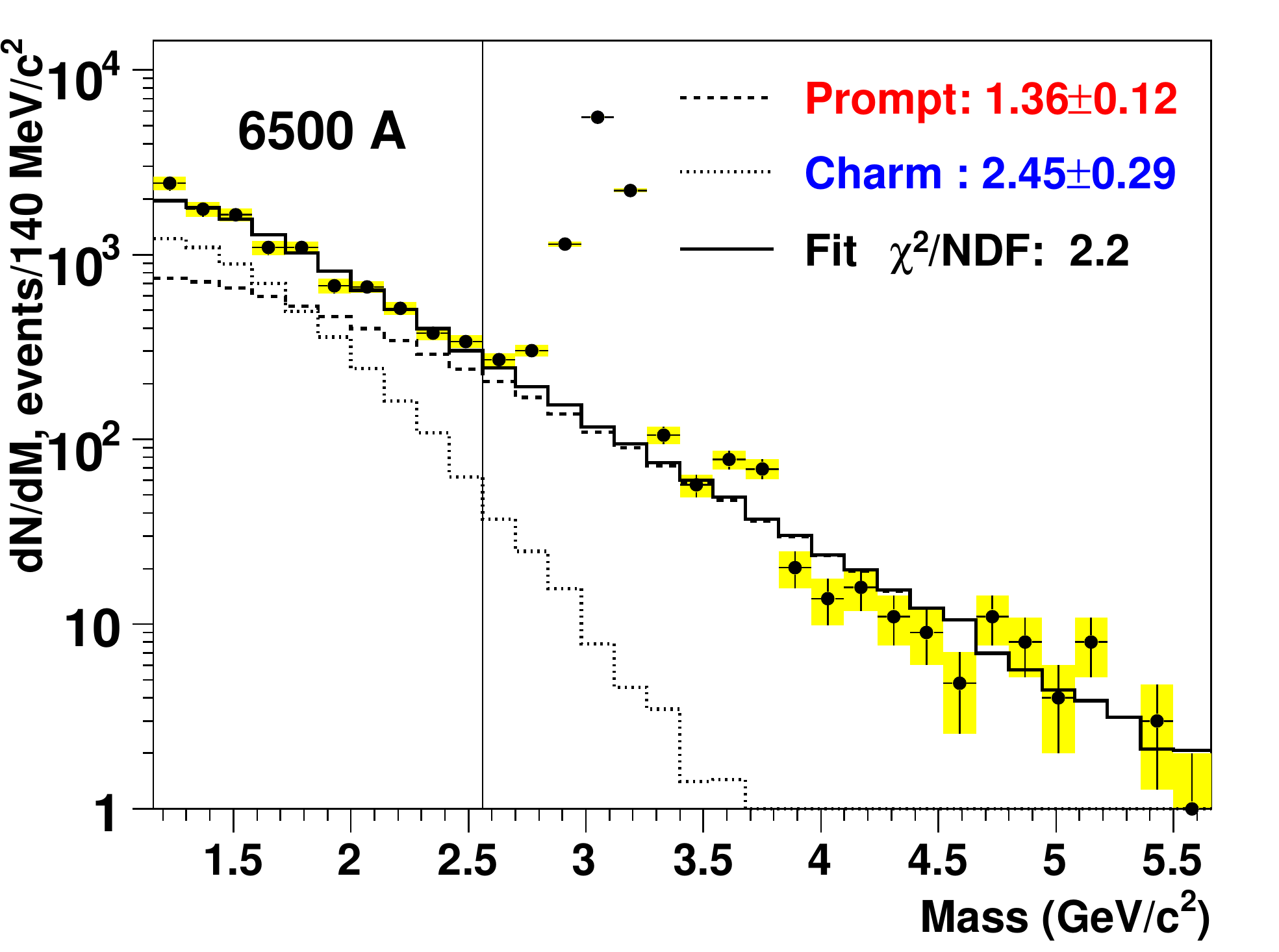}}
%
\caption{Signal dimuon mass distribution measured from the 
4000~A (\textit{top}) and 6500~A (\textit{bottom}) data samples, compared to the
superposition of Drell-Yan dimuons (dashed line) and muon
pairs from open charm decays (dotted line), scaled up with 
respect to the expected yields.}
\label{fig:massfit}
\end{figure}

Figure~\ref{fig:massfit} compares the signal dimuon mass distributions
obtained from the 4000~A and 6500~A data samples, integrated over 
collision centrality, with the sum of the Drell-Yan and 
open charm contributions.  With respect to the expected normalizations,
described in the previous paragraphs, these contributions must be 
scaled up (by the values quoted in the figure) so as to provide the best 
description of the measured signal spectrum in the dimuon mass window 
$1.16<M<2.56$~GeV/$c^2$.  Within errors, the two data samples give
perfectly compatible results.  A global fit gives scaling factors of 
$1.26\pm0.09$ for Drell-Yan and $2.61\pm0.20$ for open charm, 
with $\chi^{2}/{\rm ndf} =1.02$.
Furthermore, essentially the same numerical 
values are obtained if the analysis is redone only selecting events with a
matching $\chi^2$ below 1.5 (instead of 3).
As previously observed by NA38~\cite{NA38IMR} and NA50~\cite{NA50IMR}, 
a significant \textit{excess} of IMR muon pairs is observed, which can be well accounted 
for by increasing the charm normalization.

\begin{figure}[t]
\centering
\resizebox{0.94\columnwidth}{!}{%
\includegraphics*{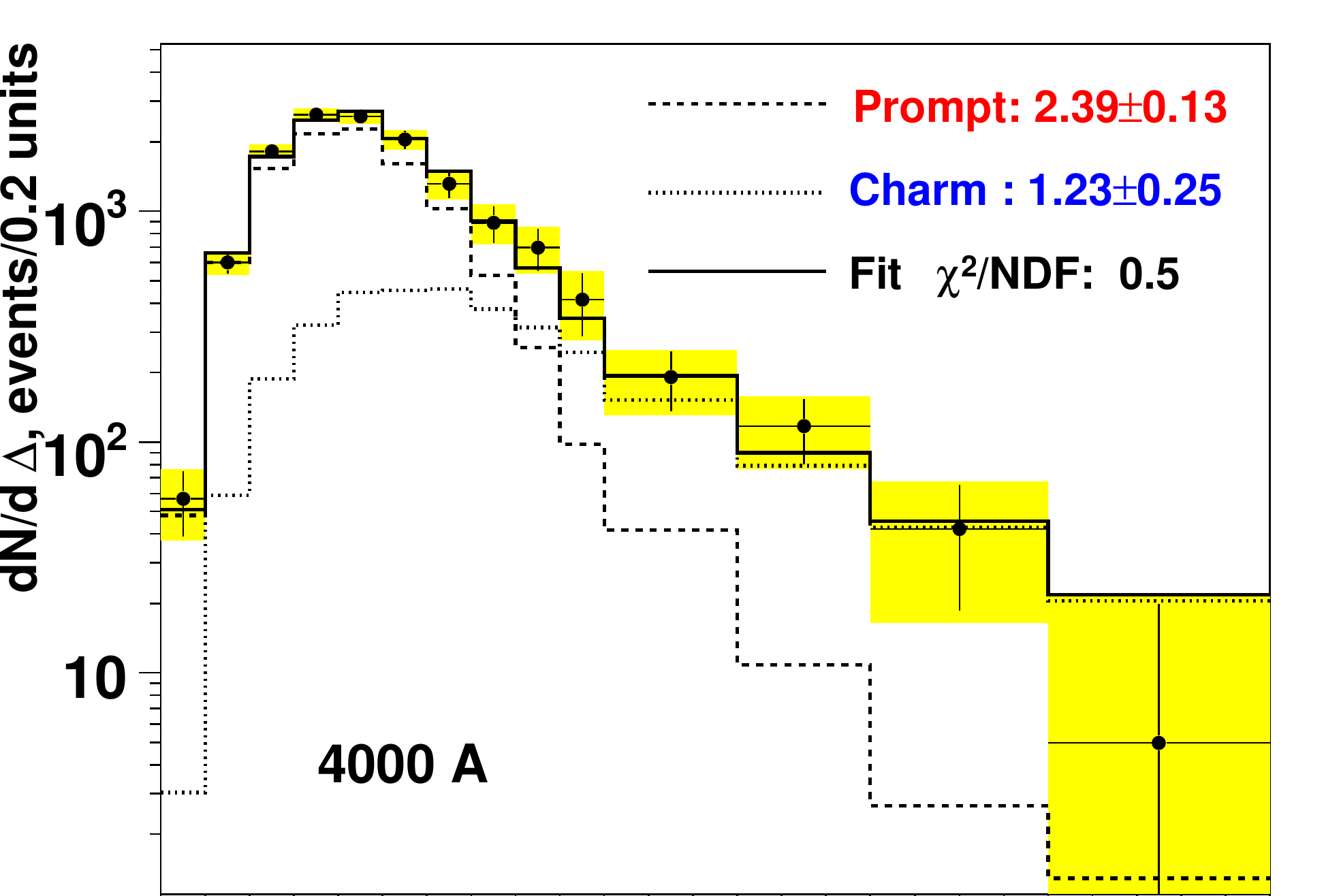}}
\resizebox{0.94\columnwidth}{!}{%
\includegraphics*{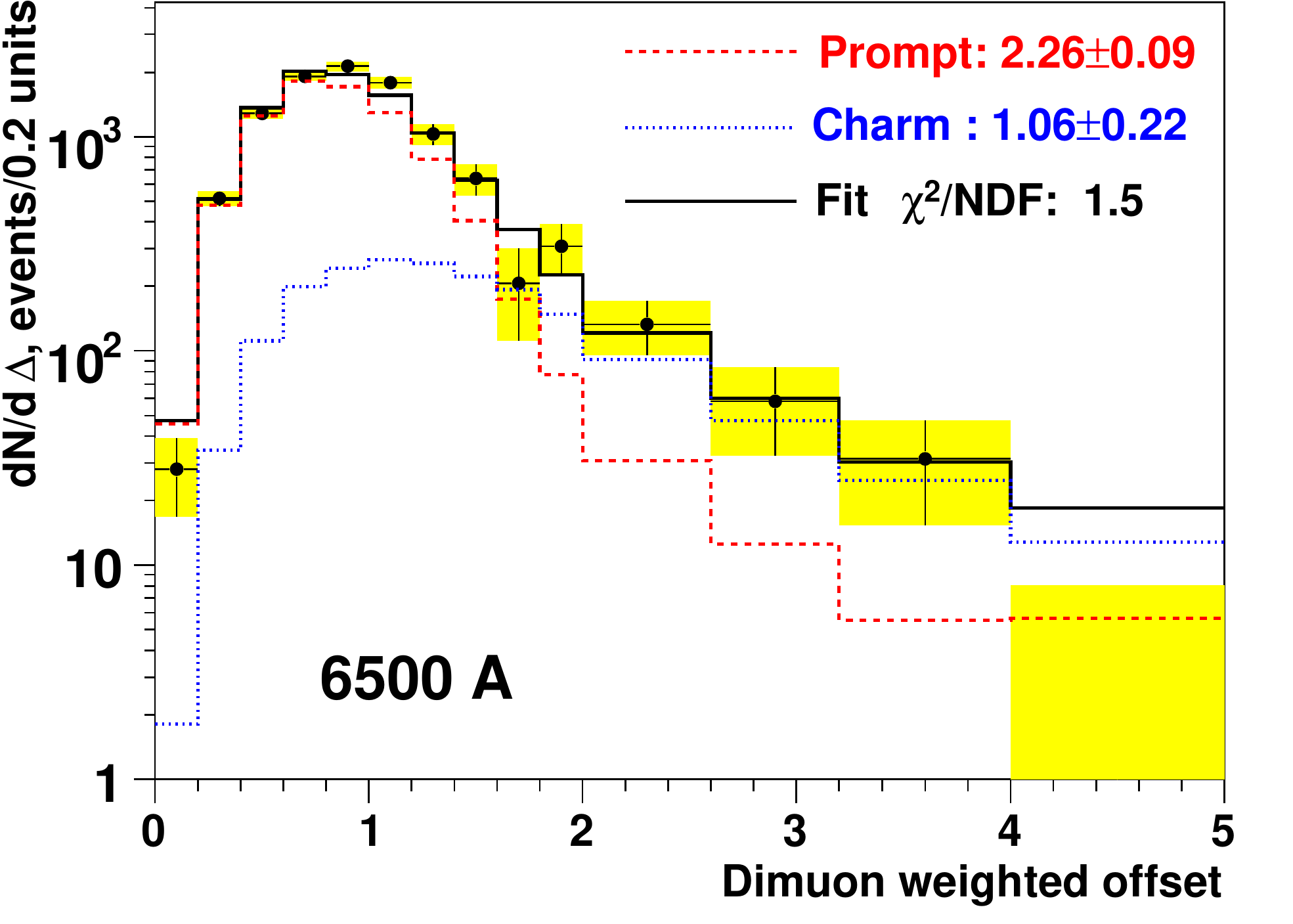}}
\caption{Same as previous figure but for the dimuon weighted offset
distributions of the dimuons in the mass range \mbox{1.16--2.56~GeV/$c^2$.}}
\label{fig:offsfit}
\end{figure}

The big advantage of NA60, with respect to the dimuon measurements
made by all other heavy-ion experiments, is the availability of the 
dimuon weighted offset variable, which provides complementary
information ideally suited to distinguish prompt dimuons from muon pairs stemming from displaced decay vertices.
Figure~\ref{fig:offsfit} shows the dimuon weighted offset distribution 
for the signal dimuons in the mass range 1.16--2.56~GeV/$c^2$,
for the 4000~A and 6500~A data samples, compared to the sum of the
two contributions: prompt dimuons and open charm decays, scaled to
provide the best fit to data.

The \emph{shape} of the prompt dimuon distribution was built
using the measured dimuons in the \jpsi\ and $\phi$
peaks, where the non-prompt signal contributions are less than 1\,\%.
The shape of the open charm distribution was defined using the 
muon pairs from the \textit{overlay Monte Carlo} simulation, including 
the additional smearing needed to reproduce the measured
 \jpsi\ and $\phi$ distributions (see \cite{ADThese}, section 8.4.1
 for details).  

The \textit{excess} dimuons are clearly concentrated in the 
region of small dimuon offsets, excluding the possibility that they are 
due to open charm decays.  The best description of the measured
distribution is obtained when the prompts contribution is scaled up by
more than a factor of two with respect to the expected Drell-Yan yield,
while the open charm contribution is compatible with the yield 
assumed by NA50 to reproduce the  IMR spectra in \mbox{p-A} collisions~\cite{NA50IMR}).
%
Fig.~\ref{fig:offsfit} underlines the fact that the dimuon weighted offset is
a really efficient discriminator between prompt and charm dimuons.
On the other hand, due to the extreme similarity of the mass
distributions of the \textit{excess} and charm dimuons
(see Fig.~\ref{fig:exc2dydd_mass}), it is clear that mass alone is completely  unable
to separate them and, therefore, perfectly accommodates the assumption
that the \textit{excess} is due to open charm, from the purely statistical
point of view (see Fig.~\ref{fig:massfit}).

A global fit of both data sets provides scaling factors of $2.29\pm0.08$ 
and $1.16\pm0.16$, for the prompt and open charm 
contributions, respectively (with $\chi^{2}/{\rm ndf} = 1.52$).  If the analysis
is redone only using events with matching $\chi^2$ below 1.5, 
instead of 3, the results remain the same within the statistical errors.
If the open charm yield is restricted to be within $10\%$ of the nominal value rather than left free in
the fit, the scaling factors become $2.43\pm0.09$ and $1.10\pm0.10$,
respectively, i.e. again the same within the statistical errors.

Since the tail of the offset distribution lacks the statistics needed to 
define the open charm normalization differentially in bins of mass,
\pt\ and centrality, its scale
is kept fixed to the value determined from the integrated sample, 
corresponding to a full phase space cross section
$\sigma_{c\bar{c}}=9.5~\mu$b/nucleon with $14\%$ statistical and
$15\%$ systematic errors accounting for the uncertainty of the expected
Drell-Yan contribution.
This is the value
needed to describe the measured data and does not depend on the 
assumed reference cross section.  
However, the extrapolation of the
cross section to full phase space depends on the 
kinematical distributions of the simulated $c$ and $\bar{c}$. The associated
uncertainty is not accounted in the quoted systematic error.
For instance, 
we would obtain a value 19\,\% smaller if we would use the conditions 
of the NA50 analysis: Pythia~5.7, different parton 
densities and without nuclear modifications, effective 
$c\bar{c} \rightarrow \mu^{+} \mu^{-}$ branching ratio of 0.97\,\% 
instead of 0.84\,\%, etc.

\subsection{Kinematic properties of the excess dimuons}

The \textit{excess} dimuons are defined as the statistical difference between the total
yield and the sum of fitted charm and nominal Drell-Yan.
Unbiased physics insight requires correcting the measured \textit{excess} for
reconstruction efficiencies and detection acceptances.
The acceptances were calculated by Monte Carlo simulations, in 
dimuon mass and \pt\ bins, for each of the two data samples,
assuming a uniform 
$\cos\theta_{\rm CS}$ distribution and the same rapidity distribution
as that of the Drell-Yan dimuons (approximately Gaussian with 
$\sigma \sim 1$).  Assuming a Gaussian rapidity distribution of width 1.5
gives 20\,\% less \textit{excess} dimuons, while a 0.5 width leads to
40\,\% more \textit{excess} dimuons.  After checking that the 4000~A and
6500~A data sets give statistically compatible results,
they were analyzed together.
In this way we obtain a two-dimensional distribution of the acceptance-corrected
\textit{excess} as a function of mass and \pt.

\begin{figure}[ht]
\centering
\resizebox{0.99\columnwidth}{!}{%
\includegraphics*{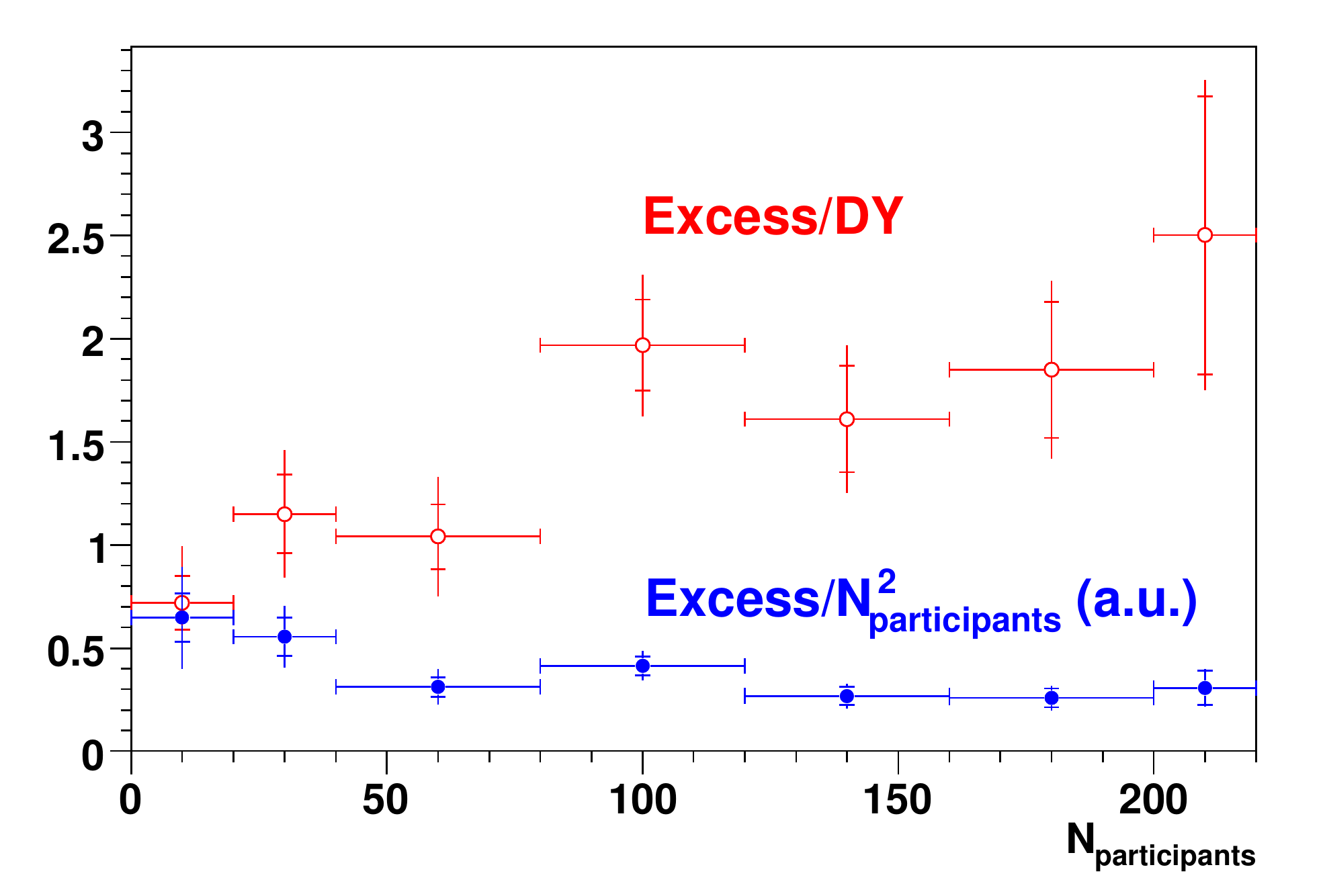}}
\caption{Ratio between the \textit{excess} and the Drell-Yan dimuon
  yields, after acceptance correction, versus centrality (open circles).
  The \textit{excess} per squared number of participants is also shown, in arbitrary units (filled
  circles).}
\label{fig:exc2dynp_mult}
\vglue -2mm
\end{figure}

Figure~\ref{fig:exc2dynp_mult} shows the 
ratio between the {\it excess} dimuon yield (corrected for acceptance 
with the assumptions just mentioned) and the expected Drell-Yan
yield (directly tak\-en from the generator), in the mass range 1.16--2.56~GeV/$c^2$ as a function of 
$N_{\rm part}$.
The smaller error bars represent the statistical errors while the 
larger ones represent the sum, in quadrature, of statistical and 
systematic uncertainties of the fitted Drell-Yan and open charm
normalizations.  We observe that the yield of 
\textit{excess} dimuons, per Drell-Yan dimuon, increases from peripheral to central In-In
collisions.  Furthermore, we see that even the most peripheral event sample
has a considerable yield of \textit{excess} dimuons (essentially identical to the 
yield of expected Drell-Yan dimuons).

To gain further insight into the properties of these \textit{excess} dimuons, we have
calculated the ratio between their yield and $N^2_{\rm part}$. The
\textit{excess}/DY was scaled  by the ratio between the number of binary 
nucleon-nucleon collisions and the squared number of participant nucleons, 
$N_{\rm coll} / N^2_{\rm part}$, calculated for each In-In
centrality bin using the Glauber model.  The resulting \textit{excess}/$N^2_{\rm part}$ 
ratio, also plotted in 
Fig.~\ref{fig:exc2dynp_mult}, shows a slight decrease with $N_{\rm  part}$.
This implies that the increase of the \textit{excess} with centrality
is somewhere in between of linear and quadratic in $N_{\rm  part}$. 
%
%
\begin{figure}[ht]
\centering
\resizebox{0.99\columnwidth}{!}{%
\includegraphics*{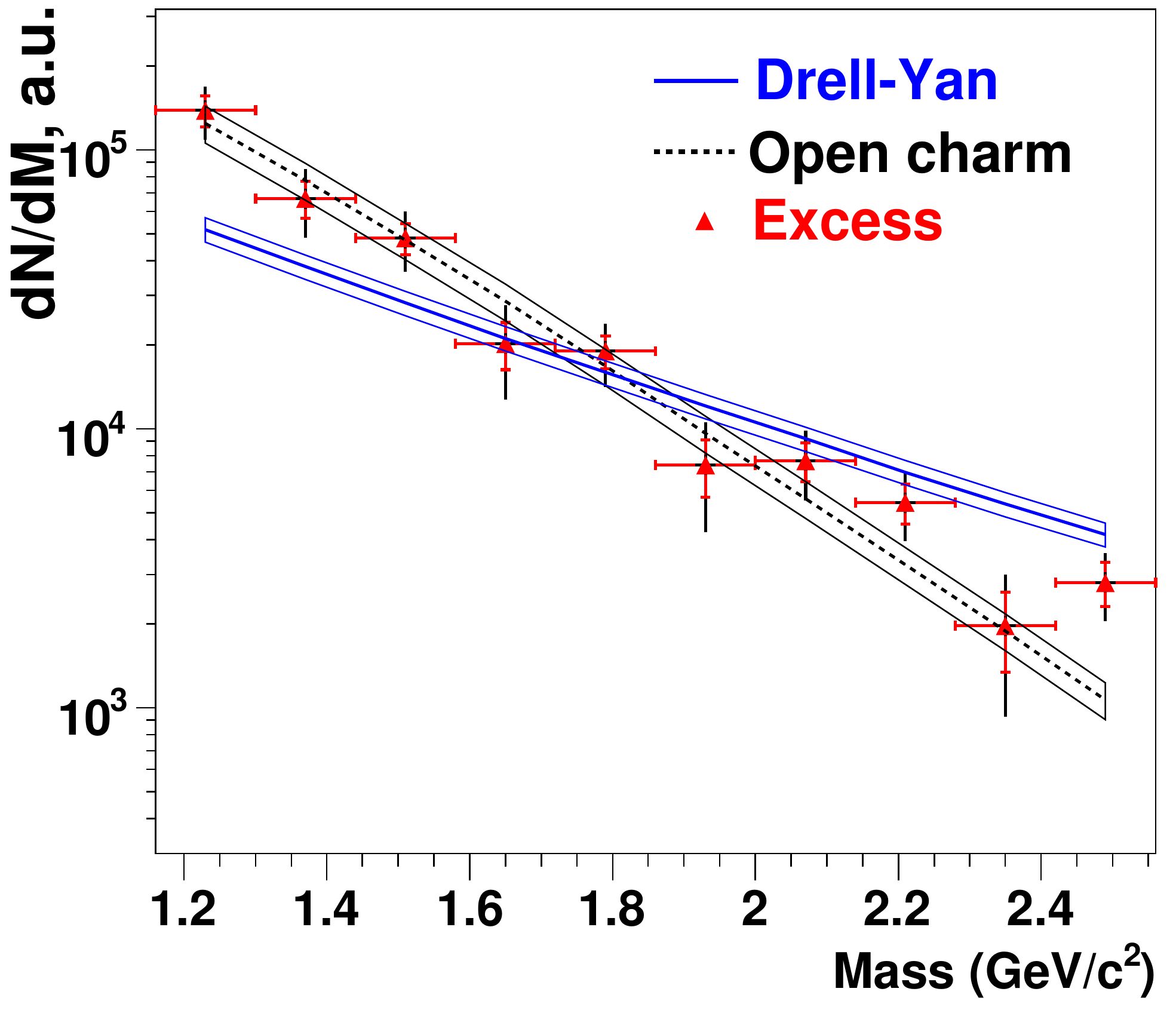}}
\caption{Mass distribution of all three contributions to the IMR spectrum, corrected for acceptances and
efficiencies as described in the text. 
From top to bottom at the highest mass bin: Drell-Yan (as expected
from the measured number of $J/\psi$ dimuons), \textit{excess} (the difference
between all prompt and Drell-Yan pairs) and open charm 
(with scaling factor of 1.16).}
\label{fig:exc2dydd_mass}
\vglue -2mm
\end{figure}

The mass distribution of the \textit{excess} dimuons, after applying the
corrections for acceptance and reconstruction efficiencies, is shown in
Fig.~\ref{fig:exc2dydd_mass}. The mass spectra of open charm and
Drell-Yan pairs are plotted for comparison. The shapes of these spectra
are directly taken from the generator. The yields reflect the fit
values discussed before; the bands correspond to the associated
systematic errors. Clearly, the mass spectrum of the \textit{excess} drops off
much more steeply with mass than that of the Drell-Yan pairs. In
contrast, the slopes are nearly the same for the \textit{excess} and open
charm. 
%
This explains why the \textit{excess}, already seen by the NA38/NA50
experiment, was found to be quite fairly described by any source,
be it prompt or delayed, with a similar mass distribution as open
charm, when using the mass spectrum only as a discriminator~\cite{CAPELLI}.

\begin{figure}[htbp]
\centering
\resizebox{0.95\columnwidth}{!}{%
\includegraphics*{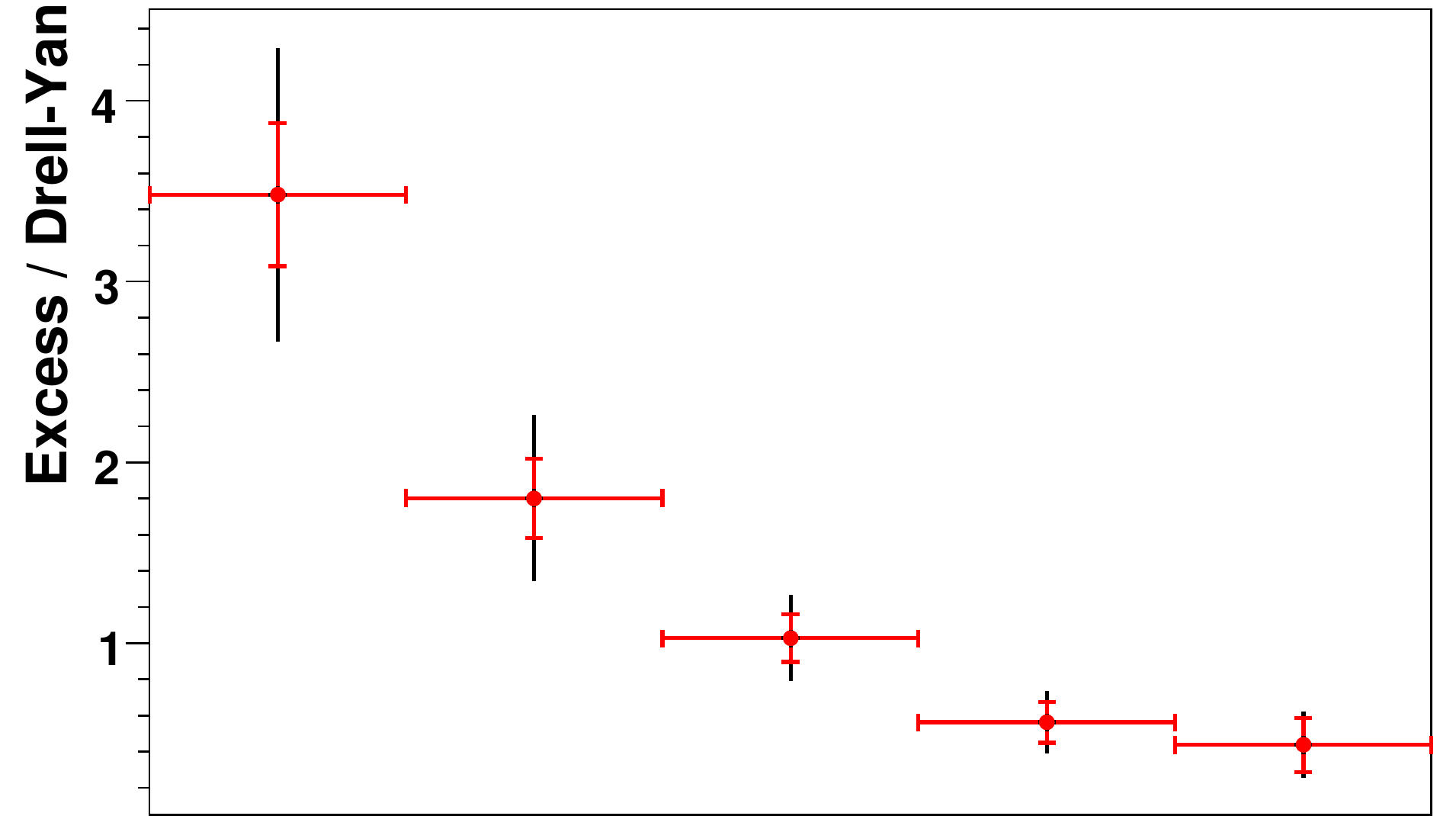}}
\vglue -0.5mm
\resizebox{0.95\columnwidth}{!}{%
\includegraphics*{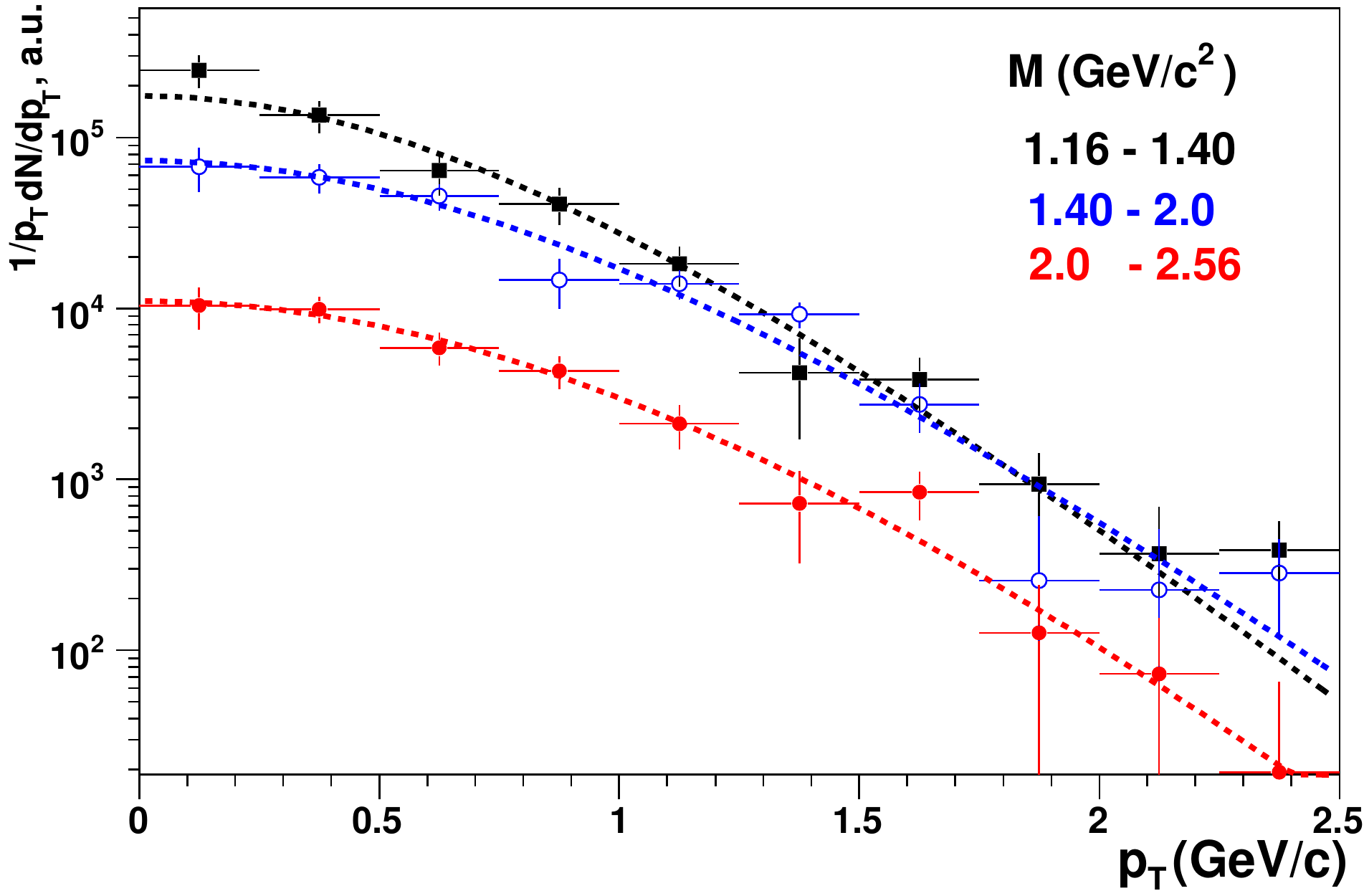}}
\resizebox{0.95\columnwidth}{!}{%
\includegraphics*{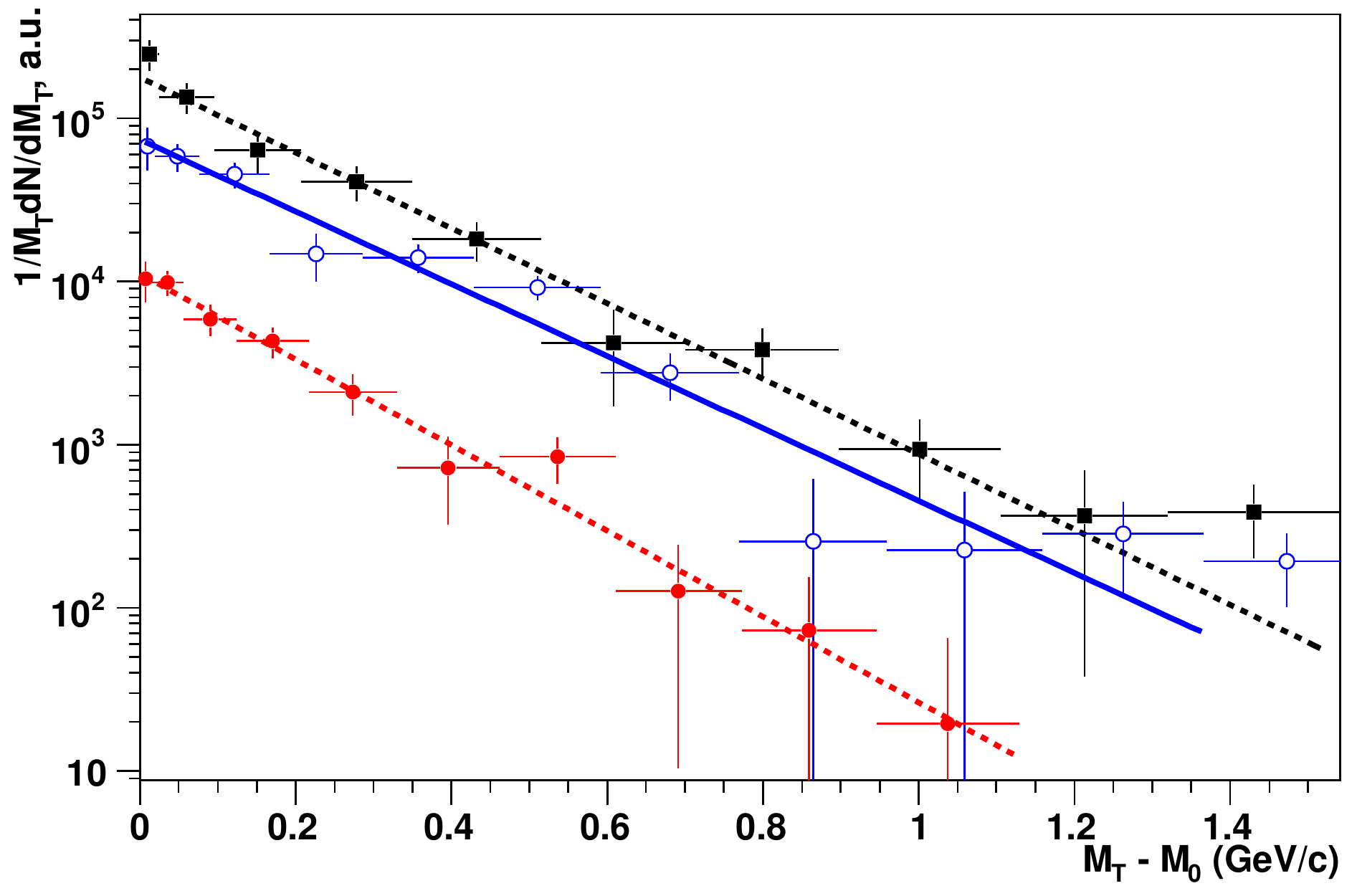}}
\caption{Ratio between the \textit{excess} and the Drell-Yan dimuon
  yields, after acceptance correction, versus dimuon transverse
  momentum (\textit{top}).
  Transverse momentum (\textit{middle}) and transverse mass
  (\textit{bottom}) spectra 
of the \textit{excess} dimuons,
in three dimuon mass windows: 1.16--1.4, 1.4--2.0 and
2.0--2.56~GeV/$c^2$.}
\label{fig:excess_pt}
\vglue -1mm
\end{figure}

Fig.~\ref{fig:excess_pt} summarizes the different aspects of the
transverse momentum distributions of the \textit{excess} dimuons. The top plot
shows the \textit{excess}/DY ratio as a function 
of the dimuon \pt, clearly indicating that the process responsible for the
production of the \textit{excess} dimuons is significantly softer than the Drell-Yan
dimuons. While in the
lowest \pt\ bin there are around 3.5 times more \textit{excess} dimuons than
expected Drell-Yan dimuons, this ratio drops to 0.5 at high \pt.
The $p_T$ spectra themselves are shown in
Fig.~\ref{fig:excess_pt}-middle, in 3 consecutive dimuon mass
windows: 1.16--1.4, 1.4--2.0 and 2.0--2.56~GeV/$c^2$. Again, a
significant deviation from the behaviour of Drell-Yan pairs is
observed. The three $p_T$ spectra are all different from each other,
while the $p_T$ spectra and the mass spectra of Drell-Yan factorize:
the primordial $k_T$=0.8~GeV/c characterizing the Gaussian
distribution is independent of mass in the region measured, i.e. from
3 to $>$10 GeV/$c^2$~\cite{Moreno90}.

Finally, Fig.~\ref{fig:excess_pt}-bottom shows the same data in terms
of dimuon transverse mass ($m_T=\sqrt{p_T^2+m^2}$) spectra, 
for the same three dimuon mass windows
defined before. All spectra are essentially exponential. However, a
steepening is observed at very low $m_T$ in the lowest mass window,
which finds its counterpart in all $m_T$ spectra observed in the low mass
region below 1~GeV/$c^2$~\cite{LMRna60}, but seems to be switched-off
in the upper two mass windows as seen here. The phenomenon is outside
any systematic errors as discussed in ~\cite{LMRna60}, but has so far
not found a convincing physical interpretation. Ignoring the low-$m_T$
rise, the data can be fit with simple exponentials 
$1/p_T~dN/dp_T = 1/m_T~dN/dm_T \sim exp(-m_T/T)$ over the complete
$p_T$-range (lines in Fig.~\ref{fig:excess_pt}-middle and bottom),
resulting in the following respective T$_{\mathrm{eff}}$ values:
$189\pm15~(stat)\pm4~(syst)$, $197\pm13\pm2$, and $166\pm17\pm4$~MeV.
The systematic errors are dominated by the uncertainties of the
Drell-Yan and open charm contributions. If the fit is instead
restricted to $p_T \ge 0.5$~GeV/$c$, consistent with~\cite{LMRna60} to
exclude the rise at low-$m_T$, the T$_{\mathrm{eff}}$ values slightly (but hardly
significantly) rise to  $199\pm 21\pm 3$, $193\pm 16\pm 2$ and $171\pm
21\pm 3$~MeV, respectively.

\section{Discussion}

The central results of the present paper are connected to two
essentially independent physics issues:
\begin{itemize}
\item The observed yield of muon pairs from D meson decays leads to $\sigma_{c\bar{c}} =
9.5\pm1.3(stat)\pm1.4(syst)~\mu$b
(the systematic error does not reflect the uncertainty related to the
kinematic distribution of the $c$ and $\bar{c}$ quarks) 
 and is compatible with the charm
production cross section deduced from the IMR dimuon data measured by
NA50 in \mbox{p-A} collisions. Charm production in \mbox{A-A}
collisions is \textit{not enhanced} relative to expectations, and it
therefore cannot be made responsible for the dimuon enhancement seen
by NA38 and NA50.
\item The dimuon enhancement in the IMR as observed before in
\mbox{Pb-Pb} collisions also exists in \mbox{In-In} collisions. It
 is now experimentally proven to be solely due to the \textit{prompt}
  component. The dimuon \textit{excess} has been 
statistically 
isolated by
  subtracting the Drell-Yan and open charm contributions from the
  total. It is on about the same level as Drell-Yan (and charm) and
  increases significantly from peripheral to central collisions
  relative to Drell-Yan and $N_{part}$. Both its mass and $p_T$
  spectra show a much steeper fall-off than Drell-Yan. Moreover, the
  $p_T$ spectra depend on mass and do not show the factorization
  between mass and $p_T$ characteristic for Drell-Yan. Conversely,
  fits to the essentially exponential $m_T$-spectra lead to inverse
  slope parameters T$_{\mathrm{eff}}$, of about 190 MeV, which do not depend on mass,
  within the (relatively large) errors.  
\end{itemize}

%

\begin{figure}[htbp]
\centering
\resizebox{0.99\columnwidth}{!}{%
\includegraphics*{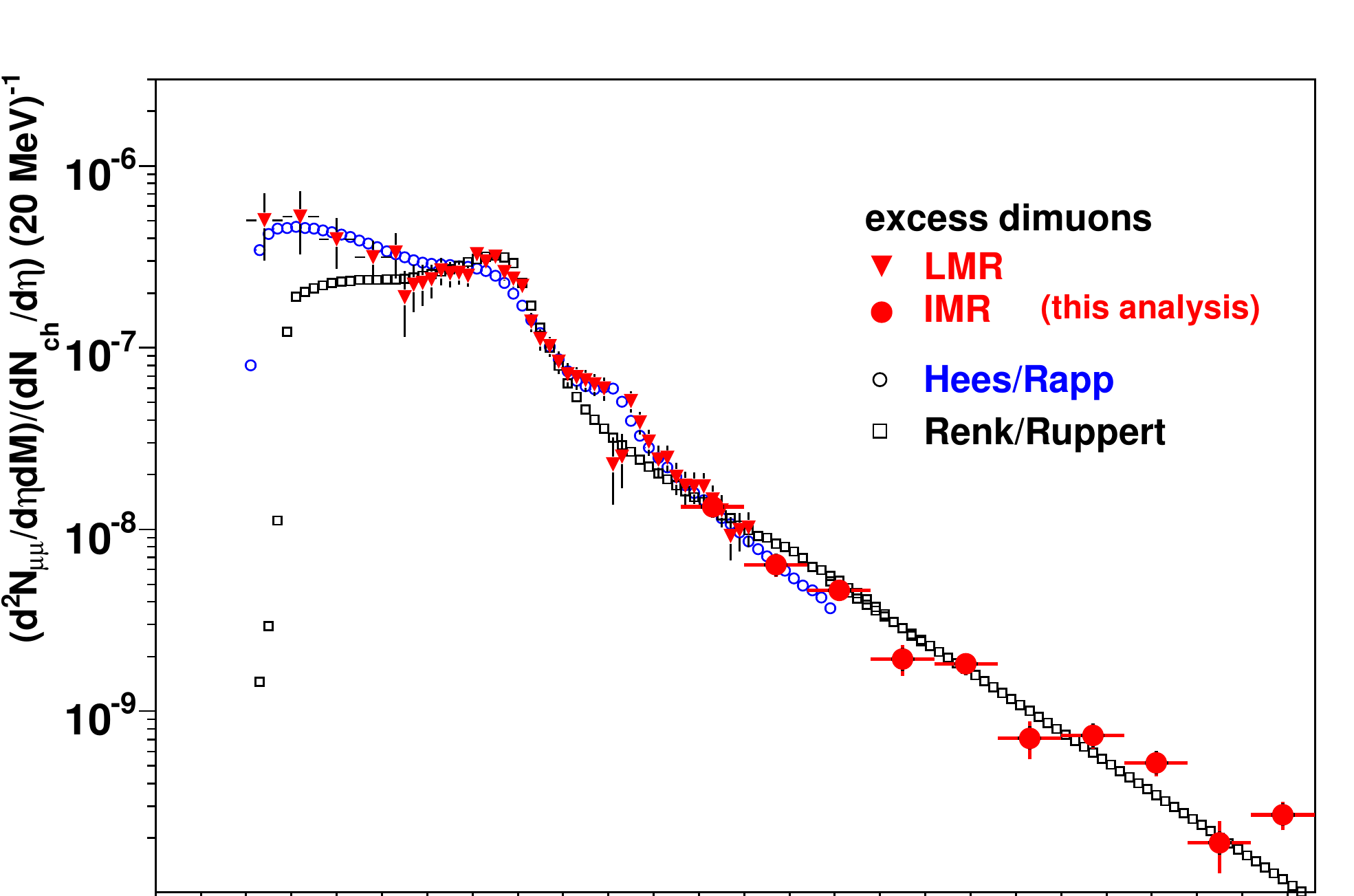}}
\resizebox{0.99\columnwidth}{!}{%
\includegraphics*{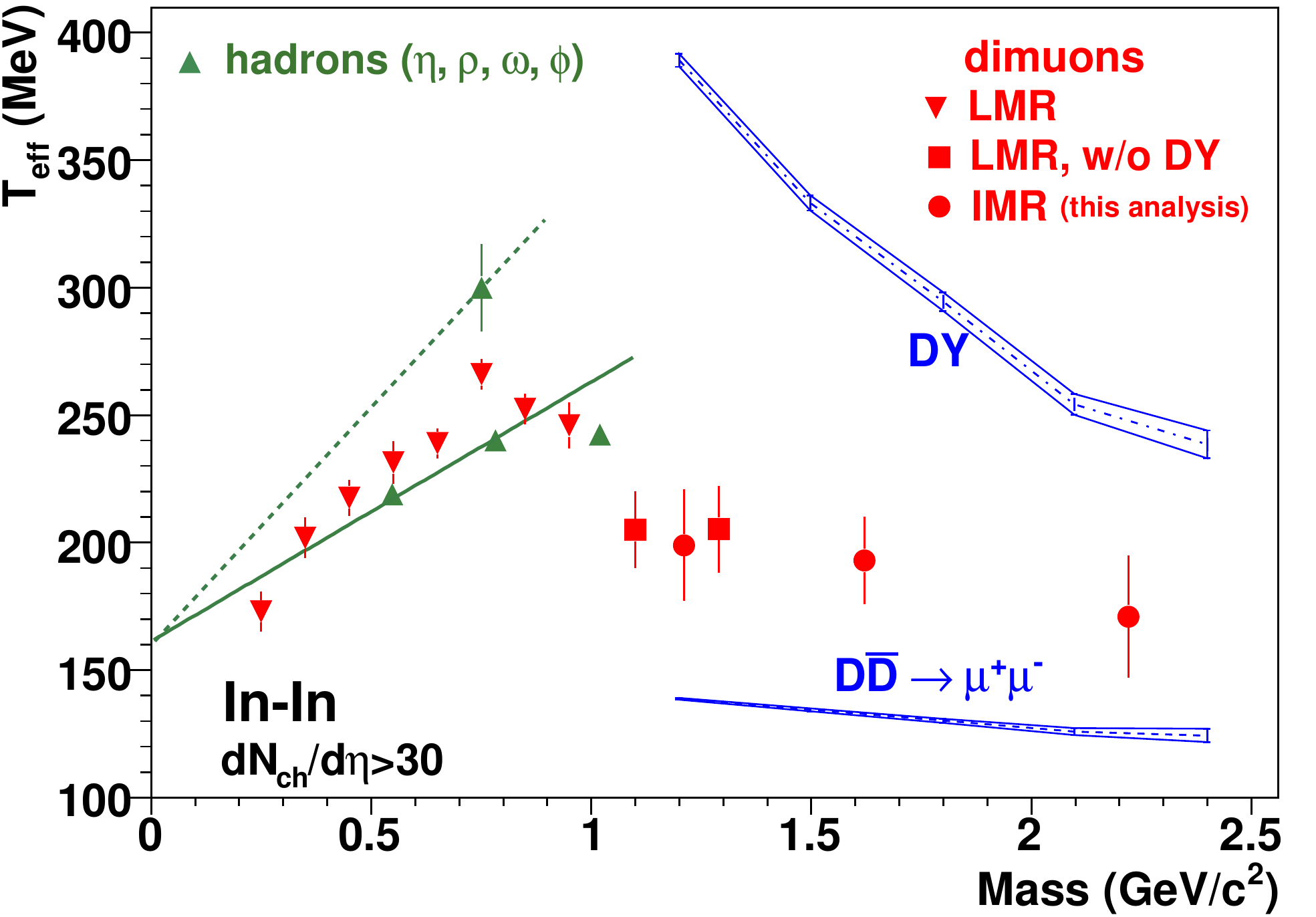}}
\caption{
\textit{Top:} Acceptance-corrected mass spectra of the excess dimuons for the
combined LMR/IMR. Errors in the LMR part are statistical; the
systematic errors are mostly smaller than that. Errors in the IMR part
are total errors. The theoretical model results are labeled
according to the authors Hees/Rapp~\cite{HEESRAPP} (EoS-B$^+$ option
is used)  and
Renk/Ruppert~\cite{RUPP}.
\textit{Bottom:} Inverse slope parameter T$_{\mathrm{eff}}$ of the
\textit{excess} $m_T$-spectra vs. dimuon mass~\cite{LMRna60,SANJAR3}. For the
LMR data M$<$1~GeV/$c^2$ (triangles), Drell-Yan is not subtracted
(would decrease the values only within the error bars~\cite{LMRna60}). The
IMR data (closed circles) correspond to the present work. 
Open charm is subtracted throughout. 
The bands show the inverse slopes for the Drell-Yan and open charm
contributions as provided by Pythia.
}
\label{fig:MTcomb}
\end{figure}

The prime contender for the interpretation of the \textit{excess} is
\textit{thermal radiation}. The remainder of this section 
places
the results reported in this paper into a wider context, relating them
to other experimental results from NA60 and to the latest theoretical
predictions on thermal radiation in the IMR.

NA60 has also studied dimuon mass and 
$p_T$
spectra in
the low mass region (LMR)
M$<$1~GeV/c$^2$~\cite{LMRna60,SANJAR3,SANJAHP08}. 
The total mass spectrum, unifying the results from the LMR analysis
and from the IMR data in Fig.~\ref{fig:exc2dydd_mass}, is shown in
Fig.~\ref{fig:MTcomb}-top.
The LMR results correspond to the integral of the $p_T$-differential
acceptance-corrected mass spectra published
previously~\cite{SANJAR3,SANJAHP08};
a cut $p_T>0.2$~GeV$/c$ is applied to avoid the very large errors in
the region of low acceptance~\cite{SANJAR3,SANJAHP08}. The mass
spectrum is absolutely normalized in the LMR region as described in
~\cite{SANJAR3}; the IMR data from Fig.~\ref{fig:exc2dydd_mass} have
independently been normalized following an analogous procedure.
Recent theoretical results on thermal radiation from two major groups
working in the field are included for comparison~\cite{HEESRAPP,RUPP},
calculated absolutely (not normalized relative to the data); a further
result~\cite{DUSLING} exists, but is left out here due to the lack of
final normalization. The general agreement between data and model
results both as to spectral shapes and to absolute yields is most
remarkable, supporting the term "thermal" used throughout this
paper. The strong rise towards low masses reflects the Boltzmann
factor, i.e. the Planck-like radiation associated with a very broad,
nearly flat spectral function. Even this part is well described
by~\cite{HEESRAPP}, due to the particularly large contribution from
baryonic interactions to the low-mass tail of the $\rho$ spectral
function in this model. Higher up in mass, the $\rho$ pole remains
visible, followed by a broad bump in the region of the $\phi$. This is
described in~\cite{HEESRAPP} as in-medium broadening of a small
fraction of the $\phi$ (caution should, however, be presently taken on
that, since corrections for 
the
resolution function of the NA60 apparatus are still under
investigation).
In the IMR region above 1~GeV/$c^2$, the description is good for both
scenarios (only available up to 1.5~GeV/$c^2$ for ~\cite{HEESRAPP}).
%

Both for the LMR and IMR, the $m_T$-spectra are pure exponentials (at
least for $p_T >$0.4~GeV/$c$), consistent with the thermal
interpretation ~\cite{RUPP,DUSLING}. Apart from the absolute scale,
they can therefore be described by one single parameter, the inverse
slope T$_{\mathrm{eff}}$ extracted from exponential fits to the data.
The combined results for T$_{\mathrm{eff}}$ 
in the LMR
and those reported in this paper are shown in
Fig.~\ref{fig:MTcomb}-bottom~\cite{LMRna60}.
For $M<1$~GeV/$c^{2}$ (triangles), 
a correction for Drell-Yan pairs is not done, due to their small
contribution~\cite{LMRna60}, the intrinsic uncertainties at low
masses~\cite{HEESRAPP} as well as the inability of Pythia to generate
Drell-Yan in this region.
The square points correspond to the extension of the LMR analysis up to 
$M=1.4$~GeV/$c^{2}$ which did not account for the systematic errors of the
Drell-Yan and open charm contributions. One should note that the square
points and circles are not statistically independent, since the two
analyses were performed on overlapping data samples. The inverse
slopes for the Drell-Yan and open charm contributions are shown
for comparison.
The difference to the excess data is most remarkable.

Below 1~GeV/c$^2$, the
inverse slope parameters T$_{\mathrm{eff}}$ are not at
all independent of dimuon mass, but monotonically rise with mass from
the dimuon threshold, where T$_{\mathrm{eff}}$ is $\sim$180~MeV, up to the nominal
pole of the $\rho$ meson, where T$_{\mathrm{eff}}$ is $\sim$250~MeV. 
This is followed by a sudden decline to
the level of T$_{\mathrm{eff}}\sim$190~MeV reported here.
That decline becomes even
more steep, jump-like, if
the slope parameters T$_{\mathrm{eff}}$ are corrected for the contribution of the
freeze-out $\rho$~\cite{SANJAHP08}. The initial rise is consistent with
the expectations for radial flow of a \textit{hadronic} source
(here $\pi^+\pi^- \rightarrow \rho$) decaying into lepton
pairs. However, extrapolating the lower-mass trend to beyond
1~GeV/c$^2$, a jump of about 50~MeV down to a low-flow situation is
extremely hard to reconcile with emission sources which continue to be
of dominantly hadronic origin in this region. Rather, the sudden loss
of flow is most naturally explained as a transition to a
qualitatively different source, implying dominantly early,
i.e. \textit{partonic} processes like $q\bar{q}\rightarrow
\mu^{+}\mu^{-}$ for which flow has not yet built up, at least at SPS
energies,
due to the "soft point" in the equation-of-state.
This may well represent the first direct, i.e. data-based
evidence for thermal radiation of partonic origin, overcoming
parton-hadron duality for the \textit{yield} description in the mass
domain (see below). The observed slope parameters T$_{\mathrm{eff}}\sim$190~MeV are
then perfectly reasonable, with a purely \textit{thermal}
interpretation without much flow, reflecting the averaging in the
space-time evolution of the fireball between the initial temperature
T$_i\sim$220-250~MeV (at the SPS) and the critical temperature T$_c\sim$170~MeV.   

Theoretically, the NA50 IMR dimuon enhancement~\cite{CAPELLI} was
successfully described as thermal radiation based on parton-hadron
duality, without specifying the individual sources~\cite{RAPPSHUR}.
However,
the same approach is not any longer appropriate for the NA60
data. The extension of the unified LMR and IMR results over the
complete M-$p_T$ plane places severe constraints on the dynamical
trajectories of the fireball evolution, allowing for more detailed
insight into the origin of the different dilepton sources on the
basis of 
radial flow, sensitive to the time ordering of the sources.
Indeed, all present scenarios ~\cite{HEESRAPP,RUPP,DUSLING} explicitly
differentiate between hadronic (mostly 4$\pi$) and partonic
contributions in the IMR. The partonic fraction ranges from 0.65 for ~\cite{HEESRAPP} 
(option EoS-B$^+$ as used in Fig.~\ref{fig:MTcomb}-top) to "dominant" in
~\cite{RUPP,DUSLING}.
However, due to remaining uncertainties in the equation-of-state, in
the fireball evolution and in the role of hard processes
~\cite{HEESRAPP}, a quantitative description of the very sensitive
inverse slope parameter T$_{\mathrm{eff}}$ in  Fig.~\ref{fig:MTcomb}-bottom is
only slowly emerging. In particular, the more recent results from the
authors of ~\cite{RUPP,DUSLING}, while very encouraging, are still
preliminary and have not yet been
formally published in their final form. A systematic comparison of
several model results to the data in Fig.~\ref{fig:MTcomb}-bottom is therefore
presently not possible. 

\section{Conclusions}

The dimuon \textit{excess} in the mass region M$>$1~GeV/c$^2$ seen in
high-energy nuclear collisions before has now been pro\-ven to be of
prompt origin. Its properties, differing from those of Drell-Yan pairs
in many ways, suggest an interpretation as thermal radiation. If
linked to supplementary information on dimuon \textit{excess} production in the
mass region $<$1~GeV/c$^2$, all indications favor an early, i.e. a
dominantly partonic emission source.
Present theoretical modelling, though still under development,
supports our interpretation.

\section{Acknowledgments}

This work was partially supported by the Funda\c{c}\~ao para a
Ci\^encia e a Tecnologia (Portugal), under the SFRH/BPD/ 5656/2001 and POCTI/FP/FNU/50173/2003 
contracts, by the Gulbenkian Foundation (Portugal) and by the
Fund Kidagan (Switzerland).

\appendix
\renewcommand{\theequation}{A-\arabic{equation}}
\setcounter{equation}{0}  
\section*{Appendix A : Combinatorial background}  
\label{appCB}

Our aim is to pick pairs of muons from different events in such a way that after applying the
trigger conditions they reproduce the observed like-sign spectra, both in shape and in absolute normalization.
The problem arises from the fact that the data used as pool for the single muon sample is already affected by the trigger conditions which induces correlations between the registered muons. We will now describe the procedure used to account for these correlations and to obtain the ``unbiased'' single muon pools.

We will denote by $P^{+}$ and $P^{-}$ the
average numbers of triggerable muons of positive and negative charge,
respectively, in a single interaction 
(the numerical values of these probabilities are always $\ll 1$ since the
probability for pion to produce a triggerable muon is $\sim 10^{-3}$).
If we neglect the correlation
induced by charge conservation between the numbers of positive and
negative hadrons (a very reasonable assumption in the case of high
multiplicity heavy-ion collisions), the number of muon pairs of
different charge combinations observed in $N$ collisions will be
\begin{eqnarray}
\label{probsq0}
N^{++} = N P^{++} = N P^{+}P^{+}/2 \nonumber \\
N^{--} = N P^{--} = N P^{-}P^{-}/2 \\
N^{+-} = N P^{+-} = N P^{+}P^{-} \quad .\nonumber
\end{eqnarray}
To account for the rejection of the same-sextant muon pairs by the
trigger, we decompose $P^{+}$ (and $P^{-}$) in the contributions from
the different sextants, $P^{+} = \sum_{i=1}^{6} {p_i^{+}}$, and
rewrite Eqs.~(\ref{probsq0}) excluding the rejected combination:
\begin{eqnarray}
\label{probtrg}
\hat{P}^{++} = \sum_{i<j}^{6} \pp_i \pp_j = \left( P^+ P^+ - \sum_{i}^{6} { {\pp_{i}}^2 } \right)\big/2  \nonumber  \\
\hat{P}^{--} = \sum_{i<j}^{6} \mm_i \mm_j =  \left( P^- P^- - \sum_{i}^{6} { {\mm_{i}}^2 } \right)\big/2 \\
\hat{P}^{+-} = \sum_{i\neq j}^{6} \pp_i \mm_j =  { P^+ P^- - \sum_{i}^{6} {\pp_{i} \mm_{i}} }\quad.  \nonumber  
\end{eqnarray}

It is worth noting that Eqs.~(\ref{probtrg}) imply: 
\begin{eqnarray}
\label{diffsqp}
\left( \hat{P}^{+-} \right)^2  -  4 \hat{P}^{++} \hat{P}^{--} = \sum_{i}^{6} { \left(P^{+} \mm_i -
P^{-} \pp_i \right) }^2- \nonumber \\
\sum_{i \neq j}^{6} {\pp_i \mm_j (\pp_i \mm_j - \pp_j \mm_i)}\quad .
\end{eqnarray}
\noindent
The well-known equation $N^{+-}=2\sqrt{N^{++}N^{--}}$ is a particular
case of Eq.~(\ref{diffsqp}) when its right-hand side vanishes. Rewriting the latter as
\begin{eqnarray}
\label{diffsqpRHS}
\sum_{i}^{6} {\left[ 
\sum_{j}^{6} { \pp_j \mm_i \left( {1 - \frac{\pp_i}{\mm_i} / \frac{\pp_j}{\mm_j} }\right)}
\right]}^2 \nonumber \\
- \sum_{i \neq j}^{6} { {\pp_i}^2 {\mm_j}^2  \left( {1 - \frac{\pp_j}{\mm_j} / \frac{\pp_i}{\mm_i} }\right) }
\quad ,
\end{eqnarray}
we see that it vanishes when the $\pp_i / \mm_i$ ratios are the same for all sextants
(in more general terms, this ratio should be constant over the whole phase space).
This is not the case in NA60 because the dipole magnet
breaks the symmetry of the azimuthal distribution of the produced
particles, in a charge dependent way.  Therefore, in order to evaluate
the {\bf CB} in NA60, we need to compute the single muon probabilities
and explicitly account for the exclusion of the same-sextant dimuons.
Since the number of same-sign pairs with muons
in the sextants $i$ and $j$ is, for the positive case, $N^{+}_{ij} = \rhop_i \rhop_j/2$, with
$\rhop_{i} = \sqrt{N} \pp_i$, we can extract $\rhop_i$ as
\begin{equation}
\label{eqalt2}
(\rhop_i)^{2} = \frac{N^{+}_{ij} N^{+}_{ik}}{N^{+}_{jk}}
\end{equation}
and then average over all possible $\{ j,k \}$ combinations.  Once the
values of $\rhop_i$ and $\rhom_i$ are known, we can determine the
fractions of positive and of negative muons, no longer being biased by
the trigger condition:
\begin{equation}
\label{eqcharge}
R^{+} = \frac{\sum{\rhop_i}}{ \sum{\rhop_i} + \sum{\rhom_i} }~~,~~~~ 
R^{-} = 1 - R^{+}\quad.
\end{equation}
We can, then, build the artificial {\bf CB} spectra according to the
following procedure:
\begin{enumerate}
\item Select randomly the charges of the two muons, according to the
probabilities given by Eqs.~(\ref{eqcharge}).
\item Select randomly the sextant of each muon, according to the
weights $\rhop_i$ and $\rhom_i$, restarting from step~1 if the two
selected sextants happen to be the same.
\item Randomly pick two muons, of charges and sextants as previously
selected, from the single muon samples built out of the measured
like-sign dimuon events.  If the two selected muons have more than one
match to \VT\ tracks ($m_1$ and $m_2$ matches, say), then we build all
possible ($m_1 \times m_2$) matched \emph{dimuons} and apply to each
of them the selection cuts applied to the measured events.
\end{enumerate}
The normalizations of these artificial {\bf CB} samples are fixed by
\begin{equation}
\label{eqcbmixnorm}
N^{+-} = \sum_{i\neq j}{\rhop_i \rhom_j}~~,~~~~
N^{++(--)} = \frac{1}{2}\sum_{i\neq j}{\rho_i^{+(-)} \rho_j^{+(-)}}\quad,
\end{equation}
\noindent
so that the generated like-sign dimuon spectra reproduce the
corresponding measured spectra.

\begin{figure}[htbp]
\centering
\resizebox{0.4\textwidth}{!}{%
\includegraphics*{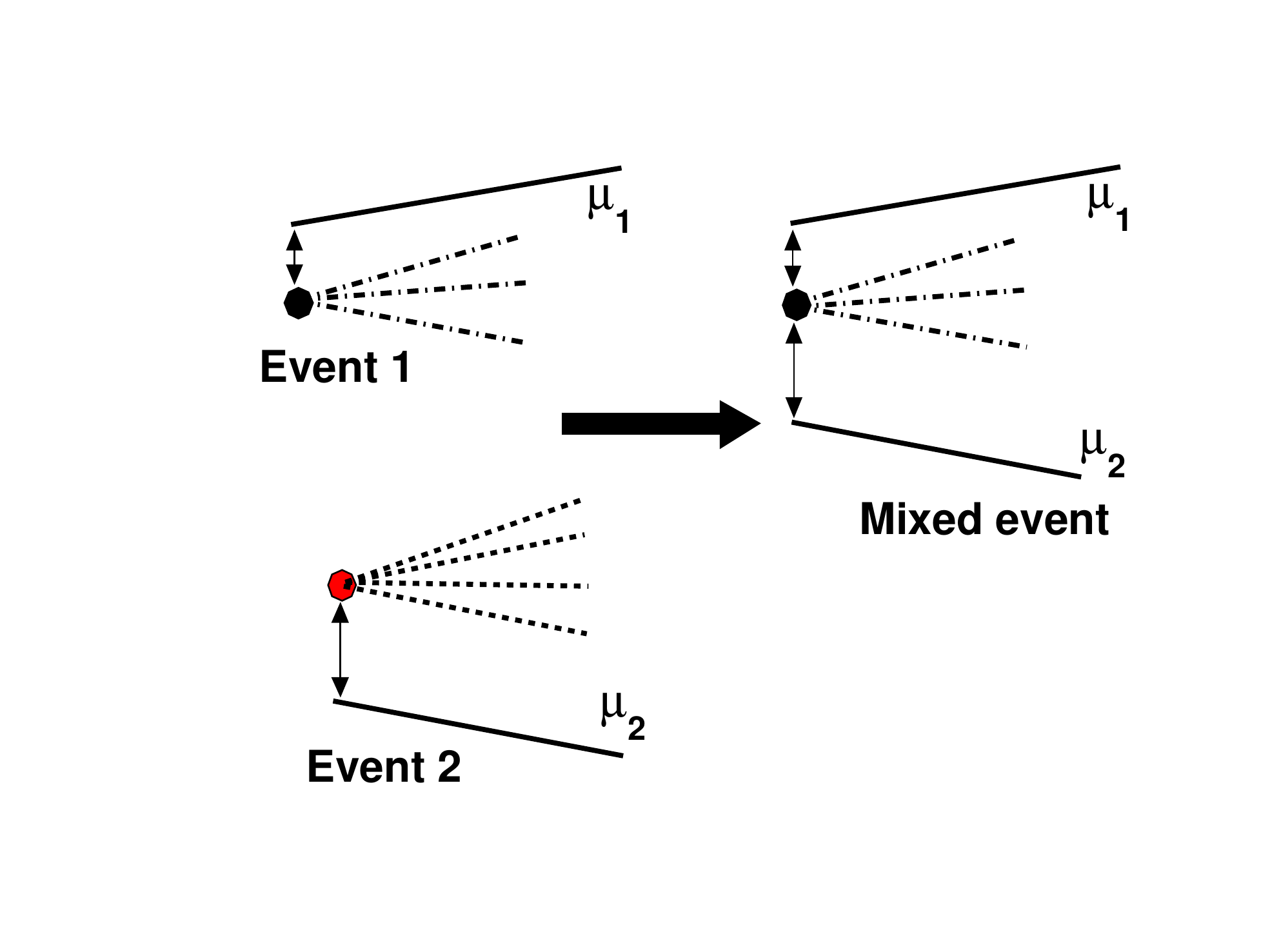}}
\caption{Schematic explanation of the method used to build the muon offsets for mixed events.}
\label{fig:moveOffs}
\end{figure}

Special care must be taken in what concerns the \textit{offsets} of
the muons in the ``mixed'' pairs.  In order to reproduce the offset
distribution of the measured combinatorial muons, 
we first randomly assign the vertex of one of the two events participating in the mixing 
to be the vertex of the generated event. Then, we modify the intercept
parameters of the muon from the other event so that with respect to this vertex it retains the 
same offset as it had in its own event, with respect to its vertex. 
 The accuracy of the
method schematically depicted in Fig.~\ref{fig:moveOffs} 
can be appreciated in Fig.~\ref{fig:lsratD}, which compares the
\textit{dimuon weighted offset} distributions of the generated and
measured like-sign muon pairs.

\renewcommand{\theequation}{B-\arabic{equation}}
\setcounter{equation}{0}  
\section*{Appendix B : Fake matches background}  
\label{appFB}

Our aim is to estimate the probability for a given muon from the \MS\ to be 
wrongly matched with the tracks in the \VT\ and to use it in building the artificial
\textit{fake dimuons} sample which reproduces both in spectral shape and in normalization
the {\bf FB} contributing to data.

Let $\epsilon$ be the probability that the correct match is present in
the set of all found matches for a muon with given kinematics (regardless
on the number of fake matches and their matching $\chi^{2}$'s).
Notice that the
correct match may be missing not only due to track
reconstruction inefficiency or the choice of the $\chi^2$ cut value, but
also because the muon was produced in a decay or interaction
downstream of the \VT, in which case its track is simply absent.
Let us denote by $\phi_n$ ($n\geq 0$) the probability for a given muon to have
$n$ fake matches in events with given characteristics such as 
interaction sub-target, multiplicity, etc. If we designate $\nu$ as the average
number of fake matches, then $\phi_n$ is Poisson distributed with mean $\nu$. 
Our derivations do not require, however, any assumption about its distribution.

These
quantities can be extracted for each muon with arbitrary precision
using a \textit{event mixing technique} in the following way: one
applies the usual matching procedure but tries to associate the \MS\ muon
of one event with the \VT\ tracks of many different events with same
characteristics (i.e.\ multiplicity, interaction sub-target, etc.).
From each such event one gets a set of \textit{a priori} fake matches
with the same $\phi_n$ and $\chi^2$ distributions as for the fakes 
in the real data. The latter distribution, $B(\chi^2)$ (scaled down by the number of
tried events per muon), after being subtracted from the $\chi^2$
distribution of the real data, provides the spectrum for the correct
matches, $S(\chi^2)$ (both  $B(\chi^2)$ and  $S(\chi^2)$ are shown in Fig.~\ref{fig:chi2match}).  
Then, the probability for a given muon to have $n$ matches (either one correct
and $n-1$ fakes or all $n$ fakes) can be written as:
%
\begin{equation}
P(n) = \epsilon \phi_{n-1} + (1-\epsilon) \phi_n
\label{eq_probn}
\end{equation}
\noindent
and the probability that the correct match is present in this $n$-plet
is
\begin{equation}
P_{pr}(n) = \epsilon \phi_{n-1} / P(n) \quad .
\label{eq_probn_corr}
\end{equation}

Provided that the correct match is present in this set of $n$ matches, Bayes' theorem states that the
fraction of times $W_{b}(n| pr)$ in which it
will have the smallest ({\it best}) $\chi^2$ is equal to the ratio of the probability
for configuration $\{\chi^2_{1,corr.}, ...\}$ to the sum of probabilities for all configurations 
with given $\chi^2$ values (i.e. $\{\chi^2_{1,fake}, \chi^2_{2,corr.}, ...\}$, etc.).
Expressing the probability of given 
arrangement of $\chi^2$ values of $n$ matches as the product of the probabilities for each $\chi^2$, 
we can write:
\begin{equation}
W_{b}(n|pr) = \frac{S(\chi^2_{1}) \prod_{j=2}^{n}{B(\chi^2_{j})} }{ \sum_{i=1}^{n}{S(\chi^2_{i}) \prod_{j\neq i}^{n}{B(\chi^2_{j})}} } 
= \frac{R_{1}}{\sum_{i=1}^{n} R_{i}}
\label{eq_probwinpres}
\end{equation}
\noindent
with $R_{i} = S(\chi^2_i)/B(\chi^2_i)$. From
Eqs.~(\ref{eq_probn}--\ref{eq_probwinpres}) we get the probability for 
the {\it best} one of $n$ matches being the correct one: 
\begin{equation}
W_{b}( n) = P_{pr}(n) W_{b}(n| pr) = \frac{ \epsilon R_{1}}{ \left( 1 + \frac{ 1-\epsilon
  }{\epsilon} \frac{\phi_n}{\phi_{n-1}} \right) \sum_{i=1}^{n} R_{i}}
  \quad .
\label{eq_probwin}
\end{equation}

It is the presence of the $\epsilon$ in Eq.~(\ref{eq_probwin}) which
makes it difficult to estimate the {\bf FB} in the \textit{best
matches} spectra using the data only: the probability of the correct
match being present in the set of the matches strongly depends on the
kinematics of the muon. Even for muons with similar kinematics
it differs for those coming from the interaction point and those originating in $\pi$ or K decays.

Consider now a pair of muons, each with its own $\epsilon_i$ and
$\phi^{(i)}_{n}$, $i=1,2$. For the sake of generality consider two extreme
possibilities: the case in which the probability of finding the correct match for the first
muon is not correlated with that of the second muon and the case in which the correct matches are
present or absent always together, with common probability
$\epsilon$. In the first case the probability
of finding $n$ and $k$ matches respectively, similarly to
Eq.~(\ref{eq_probn}), can be written as:
\begin{equation}
P(n,k)=\left[\epsilon_1 \phi^{(1)}_{n-1}+(1-\epsilon_1) \phi^{(1)}_{n}\right] \left[\epsilon_2 \phi^{(2)}_{n-1}+(1-\epsilon_2) \phi^{(2)}_{n}\right]\quad .
\label{eq_probnk}
\end{equation}
\noindent
while in the second, full correlation, case we have
\begin{equation}
P(n,k) = \epsilon \phi^{(1)}_{n-1} \phi^{(2)}_{k-1} + (1-\epsilon) \phi^{(1)}_{n} \phi^{(2)}_{k} \quad .
\label{eq_probnkcorr}
\end{equation}

Taking into account the identities $\sum_{1}^{\infty}n \phi_{n-1} =
\nu+1$, $\sum_{0}^{\infty}n \phi_{n}~=~\nu$ and $\sum_{0}^{\infty}\phi_{n}~=~1$,
the average number of matched dimuons for given pair, 
$W = \sum_{n,k=1}^{\infty} n k P(n,k)$, is equal to
\begin{equation}
W = \epsilon_1 \epsilon_2 + \left[ \epsilon_1 \nu_2 + \epsilon_2 \nu_1 + \nu_1 \nu_2  \right]
\label{eq_ndimuncorr}
\end{equation}
\noindent
for the case of absence of the correlation between the correct matches
and
\begin{equation}
W = \epsilon + \left[ \epsilon (\nu_1 + \nu_2) + \nu_1 \nu_2  \right]
\label{eq_ndimcorr}
\end{equation}
\noindent
for the correlated case. Note that in Eqs.~(\ref{eq_ndimuncorr}--\ref{eq_ndimcorr}) the 
expression in square brackets (involving the average number of fake matches per muon, $\nu$) 
gives the average number of fake dimuons which is what we want to reproduce by
the \textit{event mixing technique}.

We thus arrive at the following procedure for a given pair of muons from the
\MS\ (including those which have no matches)
\begin{enumerate}
\item For each muon of the pair we generate the ``mixed fakes'' by
selecting matches from the same number of tracks as in the event where
the pair comes from, but picking the tracks from other events with
similar characteristics.
\item Combine all mixed fakes of the first muon with all original matches
(if any) of the second one and vice-versa.  The probability of obtaining
this way $n \times k$ (including the cases of $n$ or $k=0$) \textit{a priori} fake dimuons is
\begin{equation}
F(n,k) = \phi^{(1)}_{n} \sum_{l=0}^{\infty}{ P(l,k)} + \phi^{(2)}_{k} \sum_{l=0}^{\infty}{ P(n,l) }
\label{eq_probfakedm}
\end{equation}
\noindent
which, after substitution of Eq.~(\ref{eq_probnkcorr}), leads to 
\begin{eqnarray}
F(n,k) =\phi^{(1)}_{n} \left[ \epsilon \phi^{(2)}_{k-1} + (1-\epsilon)
  \phi^{(2)}_{k} \right]  \nonumber \\
   + \phi^{(2)}_{k} \left[ \epsilon \phi^{(1)}_{n-1} + (1-\epsilon) \phi^{(1)}_{n} \right]
\label{eq_probnkcorrf}
\end{eqnarray}
\noindent
in the uncorrelated scenario and 
\begin{equation}
F(n,k) =\epsilon \left[ \phi^{(1)}_{n} \phi^{(2)}_{k-1} + \phi^{(1)}_{n-1} \phi^{(2)}_{k} \right]
\label{eq_probnkuncorrf}
\end{equation}
\noindent
for the correlated one. Averaging over all possible $\{n,k\}$
combinations (i.e.\ performing these two steps many times with
different events using the same muon pair) we get for the average
number of ``mixed fake dimuons'', $W_1$, defined as $\sum_{n,k=1}^{\infty}{n k F(n,k)}$,
\begin{eqnarray}
W_1 = \epsilon_1 \nu_2 + \epsilon_2 \nu_1 + 2 \nu_1 \nu_2 \quad ,\\
W_1 = \epsilon (\nu_1 + \nu_2) + 2 \nu_1 \nu_2 \quad ,
\label{eq_ndimfakes}
\end{eqnarray}
\noindent
for the uncorrelated and correlated cases, respectively.
\end{enumerate}

Note that these numbers reproduce the fake dimuons contribution (in
[~]) of Eqs.~(\ref{eq_ndimuncorr}) and (\ref{eq_ndimcorr}),
respectively, except for an extra factor 2 in the term corresponding
to both matches being fake.  This double counting can be easily
removed by combining the mixed fake matches of the first muon with
those of the second one and counting these dimuons with a negative
sign, thus obtaining the needed $W_2 = -\nu_1 \nu_2$ contribution.

This algorithm does not require any explicit determination of
$\epsilon$'s of $\phi_n$'s. For each pair of muons large amounts of
``mixed'' fake dimuons are generated with small weights $(W_1-W_2)/N$, where N is the
number of different events matched to the same muon pair, thus
smoothing the bin to bin fluctuations.


\end{document}